\documentclass[]{aa}  

\usepackage{graphicx}
\usepackage{txfonts}
\usepackage[colorlinks=true, citecolor=blue]{hyperref}
\usepackage{ulem}
\usepackage{caption}

\newcommand{\micron}{$\mu$m}

\newcommand{\uv}{{\it uv}}
\newcommand{\code}[1]{{\color{blue} \texttt{#1}}}
\newcommand{\pack}[1]{{\color{magenta} \texttt{#1}}}

\begin{document}

   \title{The dusty heart of Circinus}

   \subtitle{I. Imaging the circumnuclear dust in $N$-band \thanks{This work makes use of ESO Programmes 099.B-0484(A), 0104.B-0064(A), 0104.B-0127(A), 106.214U.002, and 105.205M.001.} }

   \author{J. W. Isbell \inst{1}
          \and
          K. Meisenheimer \inst{1}
          \and 
          J.-U. Pott \inst{1}
          \and
          M. Stalevski \inst{2,3}
          \and 
          K. R. W. Tristram \inst{4}
          \and 
          J. Sanchez-Bermudez \inst{1,5}
           \and 
          K.-H. Hofmann \inst{6}
          \and 
          V. Gamez Rosas \inst{7}
          \and 
          W. Jaffe \inst{7}
          \and
          L. Burtscher \inst{7}
          \and 
          J. Leftley \inst{8}
          \and 
          R. Petrov \inst{8}
          \and 
          B. Lopez \inst {8}
          \and
          T. Henning \inst{1}
          \and
          G. Weigelt \inst{6}
          \and 
          F. Allouche \inst{8}
          \and 
          P. Berio \inst{8}
          \and 
          F. Bettonvil \inst{7}
          \and
          P. Cruzalebes \inst{8}
          \and 
          C. Dominik \inst {9}
          \and 
          M. Heininger \inst{6}
          \and 
          M. Hogerheijde \inst{7,9}
          \and 
          S. Lagarde \inst{8}
          \and 
          M. Lehmitz \inst{1}
          \and
          A. Matter \inst{8}
          \and 
          A. Meilland \inst{8}
          \and 
          F. Millour \inst{8}
          \and
          S. Robbe-Dubois \inst{8}
          \and
          D. Schertl \inst{6}
          \and 
          R. van Boekel \inst{1}
          \and
          J. Varga \inst{7}
          \and 
          J. Woillez \inst{10} 
          }

   \institute{Max-Planck-Institut f\"ur Astronomie (MPIA), K\"onigstuhl 17, 69117 Heidelberg, Germany\\
              \email{isbell@mpia-hd.mpg.de}
      \and Astronomical Observatory, Volgina 7, 11060 Belgrade, Serbia
      \and 
      Sterrenkundig Observatorium, Universiteit Gent, Krijgslaan 281-S9, Gent, 9000, Belgium
      \and 
      European Southern Observatory, Alonso de C\'ordova 3107, Vitacura, Santiago, Chile
      \and 
      Instituto de Astronom\'ia, Universidad Nacional Aut\'onoma de M\'exico, Apdo. Postal 70264, Ciudad de M\'exico 04510, Mexico
      \and 
      Max-Planck-Institut f\"ur Radioastronomie, Auf dem H\"ugel 69, D-53121 Bonn, Germany
      \and
      Leiden Observatory, Leiden University, Niels Bohrweg 2, NL-2333 CA Leiden, The Netherlands
      \and 
      Laboratoire Lagrange, Universit\'e C\^ote d'Azur, Observatoire de la C\^ote d'Azur, CNRS, Boulevard de l'Observatoire, CS 34229, 06304 Nice Cedex 4, France
      \and 
      Anton Pannekoek Institute for Astronomy, University of Amsterdam, Science Park 904, 1090 GE Amsterdam, The Netherlands
      \and
      European Southern Observatory Headquarters, Karl-Schwarzschild-Stra\ss e 2, 85748 Garching bei M\"unchen, Germany
      }

   \date{Received February 7, 2022; accepted April 27, 2022}

 
  \abstract
   {Active galactic nuclei play a key role in the evolution of galaxies, but their inner workings and physical connection to the host are poorly understood due to a lack of angular resolution. Infrared interferometry makes it possible to resolve the circumnuclear dust in the nearby Seyfert 2 galaxy, the Circinus Galaxy. Previous observations have revealed complex structures and polar dust emission but interpretation was limited to simple models. The new Multi AperTure mid-Infrared Spectro-Scopic Experiment (MATISSE) makes it possible to image these structures for the first time.}
   {We aim to precisely map the morphology and temperature of the dust surrounding the supermassive black hole through interferometric imaging.}
   {We observed the Circinus Galaxy with MATISSE at the Very Large Telescope Interferometer (VLTI), producing 150 correlated flux spectra and 100 closure phase spectra. The novel inclusion of closure phases makes interferometric imaging possible for the first time. We reconstructed images in the $N$-band at $\sim 10$ mas resolution. We fit blackbody functions with dust extinction to several aperture-extracted fluxes from the images to produce a temperature distribution of central dusty structures.}
   {We find significant substructure in the circumnuclear dust: central unresolved flux of $\sim0.5$ Jy, a thin disk 1.9 pc in diameter oriented along $\sim 45^{\circ}$, and a $\sim 4 \times$ 1.5 pc polar emission extending orthogonal to the disk. The polar emission exhibits patchiness, which we attribute to clumpy dust. Flux enhancements to the east and west of the disk are seen for the first time. We distinguish the temperature profiles of the disk and of the polar emission: the disk shows a steep temperature gradient indicative of denser material; the polar profile is flatter, indicating clumpiness and/or lower dust density. The unresolved flux is fitted with a high temperature, $\sim 370$ K. The polar dust remains warm ($\sim 200$ K) out to 1.5 pc from the disk. We attribute approximately $60$\% of the 12 $\mu$m flux to the polar dust, 10\% to the disk, and 6\% is unresolved; the remaining flux was resolved out. The recovered morphology and temperature distribution resembles modeling of accretion disks with radiation-driven winds at large scales, but we placed new constraints on the subparsec dust.}
   {The spatially resolved subparsec features imaged here place new constraints on the physical modeling of circumnuclear dust in active galaxies; we show strong evidence that the polar emission consists of dust clumps or filaments. The dynamics of the structures and their role in the Unified Model remain to be explored.}


   \maketitle
%

\section{Introduction}
Active galactic nuclei (AGN) are thought to play a crucial role in the formation and evolution of its host galaxy. Moreover, understanding the dust in the vicinity of supermassive black holes is key to understanding how AGN are fed and how they interact with their hosts. The dust traces dense molecular gas which feeds the AGN. Large, obscuring dusty structures are thought to be responsible for both funneling material toward the central engine, and for distinguishing between Seyfert 1 and Seyfert 2 AGN. In the original Unified Model of AGN \citep{antonucci1993, urry1995,netzer2015}, a central obscuring torus of dust is oriented such that the broad-line region of the AGN is directly visible (Seyfert 1) or such that its observation is blocked by the torus (Seyfert 2; hereafter Sy2). So in order to fully understand the accretion process and the life cycle of an AGN, one must understand the parsec-scale dust structures surrounding it.

The so-called torus is comprised of several key features which vary in temperature from $<200$ K to $1500$ K and scale from tenths of a parsec to hundreds of parsecs.
The inner edge is the radius at which radiation from the accretion disk (AD) causes the dust to sublimate. The sublimation radius is dependent on both the luminosity of the AD and the dust composition, but typically $\sim 0.1$ pc for dust evaporating at $1500$ K, for a $L \sim 1 \times 10^{10} L_{\odot}$ AGN. 
Beyond the sublimation zone, it is thought that a dense disk or torus of material is responsible for both hiding the broad line region (BLR) in Sy2 AGN and for feeding the AD. Previous mid-infrared (MIR) interferometric studies revealed that many ``tori'' have an additional component in the form of a polar extension \citep[see, e.g.,][]{honig2012,burtscher2013,lopez-gonzaga2016,leftley2018}, the Circinus Galaxy's chief among them \citep[][]{tristram2007,tristram2014}. The polar component is thought to be a radiation-driven outflow \citep[e.g.,][]{wada2012,wada2016}, and it can represent a key mechanism of AGN feedback. This is called the fountain model, and it was shown by \citet{schartmann2014} to reproduce the MIR polar extension and dusty hollow cone in the Circinus Galaxy (hereafter Circinus). 
A key finding of spectral energy distribution (SED) fits to nearby AGN as well as comparisons to radiative transfer models is that the dust in the central structures (and particularly in the wind) must be clumpy, allowing dust to reach high temperatures and exhibit ``blue'' spectra even at large distances from the AD \citep[][]{krolik1988,nenkova2008a, honig2017,martinez-paredes2020,isbell2021}. 
The exact nature of these components and how they are connected to each other and to the host galaxy remains an open question.  A holistic model of the central dust distribution is shown in \citet{izumi2018}, but only the resolution offered by infrared interferometry can probe the subparsec details of the dust near the active nucleus.

The Multi AperTure mid-Infrared Spectro-Scopic Experiment (MATISSE) is the second-generation MIR interferometer on the Very Large Telescope Interferometer (VLTI) at the European Southern Observatory (ESO) Paranal site \citep{lopez2014,lopez2022}. MATISSE combines the light from four unit telescopes (UTs) or four auxiliary telescopes (ATs) measuring six baselines in the \textit{L}-, \textit{M}-, and \textit{N}-bands simultaneously. MATISSE furthermore introduces closure phases to MIR inteferometry. The combination of the phase measurements on any three baselines $\phi_{ijk} \equiv \phi_{ij} + \phi_{jk} - \phi_{ik}$ is called the closure phase; this summation cancels out any atmospheric or baseline-dependent phase errors \citep[][]{jennison1958, monnier2003}. Closure phases are crucial for imaging because they probe the spatial distribution of target flux and because they are unaffected by atmospheric turbulence. Recent imaging studies of NGC\,1068 with GRAVITY \citep{gravitycollaboration2020} and MATISSE \citep{gamezrosas2022} have illustrated the power of this approach in revealing new morphological details and spatially resolved temperature measurements of the circumnuclear dust. Until this work, NGC\,1068 was the only AGN to have been imaged with MATISSE.

Circinus is of particular interest as it is one of the closest Sy2 galaxies \citep[at a distance of 4.2 Mpc][]{freeman1977, tully2009} and the second brightest in the MIR (only fainter than NGC 1068). 
Circinus is a prototypical Sy2 galaxy, exhibiting narrow emission lines \citep[][]{oliva1994,moorwood1996} and an obscured broad-line region \citep[BLR; ][]{oliva1998}, as well as bipolar radio lobes \citep[][]{elmouttie1998} and an optical ionization cone \citep[][]{marconi1994,maiolino2000,wilson2000,mingozzi2019}. Additionally, Circinus exhibits a Compton thick nucleus and a reflection component in X-rays \citep[][]{matt1996,smith2001,soldi2005,yang2009}. Finally, in- and outflows and spiral arms have been observed in CO down to $\sim5$ pc scales \citep[][]{curran1998, izumi2018,tristram2022}, further indicating the complexity of the central structures.

Circinus was observed extensively with the first generation MIR interferometer, the MID-infrared Interferometric instrument \citep[MIDI;][]{leinert2003}, in the $N$-band \citep[e.g.,][hereafter T14]{tristram2007, tristram2014}. These observations showed a warm ($\sim 300$ K) dust disk roughly aligned with the water maser emission \citep{greenhill2003}, but the flux was dominated by large scale ($\gtrsim 100$ mas) emission roughly orthogonal to the disk. The orientation of the large scale emission's major axis was found to differ significantly from the optical ionization cone central angle (PA$_{\rm opt.}$ = $-45^{\circ}$ vs PA$_{\rm dust} = -73^{\circ}$), and follow-up modeling work by \citet{stalevski2017,stalevski2019} has indicated that the polar-extended dust emission may come from an edge-brightened outflow cone.   

The proximity and declination of Circinus (at around $-60^{\circ}$) make it an ideal target for imaging with MATISSE, as it provides high spatial resolution (10 mas $=$ 0.2 pc) and because its nearly circular $uv$-tracks aid in the production of high fidelity reconstructions. 
MATISSE provides the first MIR measurements of the closure phase, which sample the (a)symmetry of a source and are crucial for image reconstruction. Previous analysis relied on Gaussian model fitting, which is a smooth, simplified representation of the source emission; but interferometric image reconstruction has the potential to build on these results through model-independent sampling of the source structure. Herein we present the first image reconstructions of the $N$-band circumnuclear dust in Circinus.

This paper is organized as such: in \S \ref{sec:obs} we present the observations entering this work as well as the data reduction methods. In \S \ref{sec:imarec} we lay out the interferometric image reconstruction process and final image reconstruction parameters. We also compute image errors and assess the morphology of the resulting structure. In \S \ref{sec:temperature} we measure the temperature distribution of the dust in the central structure via blackbody fitting. In \S \ref{sec:discussion} we analyze the various components of the central dust structure in Circinus and discuss their implications. Finally, we conclude and summarize in \S \ref{sec:conc}.

\begin{table}[]
    \centering
    \caption{VLTI/MATISSE observations entering this analysis}
    \small
    \begin{tabular}{ll|ccc}
    TPL Start & Target & N$_{cycles}$& $\tau_0$ [ms] & Seeing ["]  \\ \hline\hline 
    2020-03-13T04:02:11 & Circinus & 2 & 6.4 & 0.73 \\
    2020-03-13T04:56:22 & Circinus & 1 & 7.1 & 0.58 \\
    2020-03-14T03:53:00 & Circinus & 2 & 7.1 & 0.65 \\
    2020-03-14T04:31:58 & Circinus & 2& 4.9 & 0.91 \\
    2020-03-14T04:51:12 & Circinus & 4& 7.3 & 0.63 \\
    2020-03-14T07:57:12 & Circinus & 2 & 6.6 & 0.54 \\
    2020-03-14T08:57:48 & Circinus & 4 & 8.0 & 0.47 \\
    2021-02-28T06:32:19 & Circinus & 2& 10.8 & 0.79 \\
    2021-02-28T07:42:00 & Circinus & 2& 8.8 & 0.81 \\
    2021-06-01T03:10:17 & Circinus & 2& 4.7 & 0.70 \\
    2021-06-01T04:29:41 & Circinus & 2& 5.8 & 0.54 \\ \hline
    Calibrators & & & \\ \hline
    2020-03-13T04:40:24 & HD120404 &1& 6.0 & 0.56 \\
    2020-03-14T05:59:29 & HD120404 &1& 7.9 & 0.48 \\
    2020-03-14T08:31:10 & HD120404 &1& 7.4 & 0.55 \\
    2021-02-28T06:18:58 & HD120913 &1& 9.0 & 0.79 \\
    2021-02-28T07:07:46 & HD120404 &1& 5.8 & 1.06 \\
    2021-06-01T02:38:46 & HD119164 &1& 5.2 & 0.78 \\
    2021-06-01T03:59:25 & HD120404 &1& 6.2 & 0.47 \\
    \end{tabular}
    \tablefoot{Seeing and coherence time ($\tau_0$) values are given from the start of each observing block; N$_{cycles}$ is the number of observed interferometric cycles, consisting each of four 1\,min long exposures with changing configurations of the BCD.} 
    \label{tab:tab1}
\end{table}
\section{Observations and data reduction}
\label{sec:obs}
\subsection{MATISSE observations}
\begin{figure}
    \centering
    \includegraphics[width=.5\textwidth]{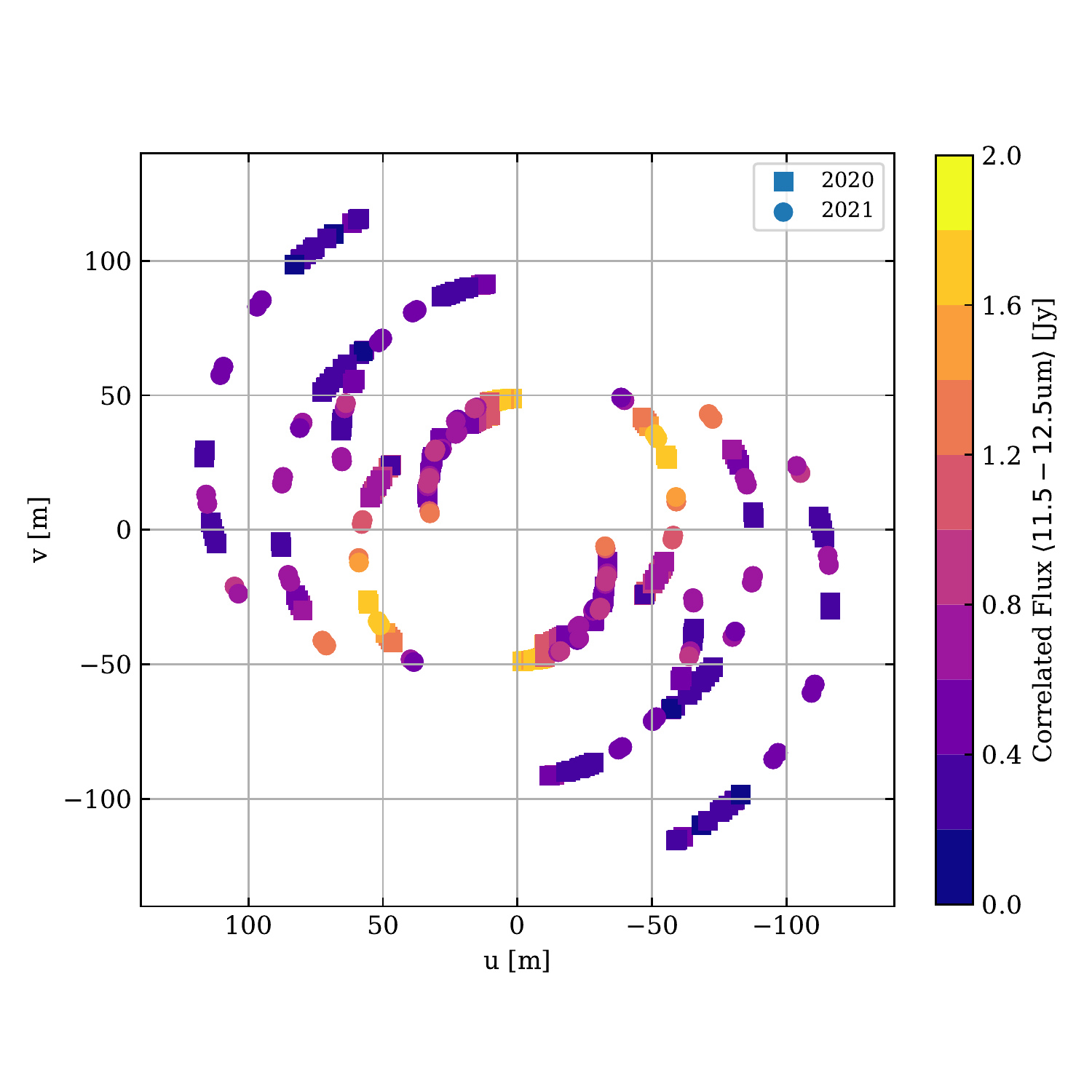}
    \caption{MATISSE \uv-coverage from all 25 exposure cycles. Squares denote observations taken in 2020, while circles represent observations from 2021. The mean correlated flux between 11.5 and 12.5 \micron~is used as the color scale. The discrete color binning is done in 0.2 Jy intervals, based on the measured correlated flux uncertainties. }
    \label{fig:uv}
\end{figure}

The MATISSE observations of Circinus were carried out on 13--14 March 2020, 27 Feb 2021, and 31 May 2021 as part of guaranteed time observations. Data were taken with low spectral resolution in both the $LM$- (3--5 $\mu$m) and $N$-bands (8--13 $\mu$m). The observations were taken using the unit telescope (UT) configuration, with physical baselines ranging from 30 m to 140 m. At 12 \micron~this corresponds to angular resolutions between 9 and 41 mas with a ``primary beam'' of 153 mas. Each observation sequence consists of two sky exposures, a number of exposure cycles, N$_{cycles}$, consisting each of four 1\,min interferometric exposures with different configurations of the beam commuting device (BCD) of MATISSE, as well as optional photometric exposures while chopping \citep[for details see][]{lopez2022}. 
Near the end of the night of 14 March 2020, we opted to repeat more exposure cycles to reduce the overhead time of re-acquisition on the target. The exact number of exposure cycles, along with the atmospheric conditions at the start of each observation, are given in Table 1. The observing conditions on 14 March 2020 were excellent, while on 13 March 2020 high-altitude cirrus negatively impacted acquisition, guiding, and adaptive optics in several individual exposures; we note that the final correlated flux error estimates on this night are higher. Observations on 28 Feb and 01 Jun 2021 were unaffected by such issues. We show the combined \uv-coverage of all the observations in Fig. \ref{fig:uv}.

On each night, we observed the calibration star HD120404 ($F_{12\mu{\rm m}} = 13$ Jy) directly before and/or after the Circinus observations. The atmospheric conditions at the start of each calibrator observation are given in Table 1. This star serves a spectral calibrator, an instrumental phase calibrator, and an instrumental visibility calibrator. It has a MIR spectrum given by \citet{vanboekel2004}, and its diameter is given as 2.958 mas in \citet{cruzalebes2019}. During the Feb. and May 2021 observations, we observed secondary calibrators, HD120913 ($F_{12\mu{\rm m}}=5.7$ Jy) and HD119164 ($F_{12\mu{\rm m}}=1.2$ Jy) in order to perform cross-calibration and closure phases accuracy checks. 

We focus hereafter solely on the $N$-band observations. While $LM$-band data were recorded simultaneously, the low total flux of the core \citep[$458.16 \pm 39.18$ mJy;][]{isbell2021} results in very faint correlated fluxes for even a marginally resolved source. We leave the analysis of the $LM$ data for a future paper, awaiting improvements in low-signal-to-noise calibration.

\subsection{MATISSE data reduction and calibration}
The $N$-band data were initially reduced using the MATISSE data reduction software\footnote{\href{https://www.eso.org/sci/software/pipelines/matisse/}{https://www.eso.org/sci/software/pipelines/matisse/}} (DRS) version 1.5.1. We used the coherent reduction flags \code{corrFlux=TRUE} and \code{coherentAlgo=2} in order to produce correlated fluxes using the coherent integration algorithm as employed in the MIDI Expert Work Station \citep[EWS; ][]{jaffe2004b} and used in T14.
We also use spectral binning 21 px (= 1 \micron) and the default values for all other parameters.

The correlated flux, $F(u,v,\lambda)$ was then calibrated in the standard way: 
\begin{equation}
F_{\rm targ}^{\rm cal}(u,v,\lambda) = F_{\rm targ}^{\rm raw}(u,v, \lambda) / F_{\star}^{\rm raw}(u,v, \lambda) \times F_{\rm \star}^{\rm tot}(u=0,v=0,\lambda), 
\end{equation}
where $F_{\star}^{\rm raw}$ is the raw flux (in counts) of the calibrator, $F_{\star}^{\rm tot}$ is the catalog flux of the calibrator, and $F_{\rm targ}^{\rm raw}$ is the raw flux of the target. This assumes the calibrator is unresolved; for the selected calibrators with diameter $< 3$ mas this is the case.

Within an observing cycle, individual exposures are taken minutes apart.  
The standard deviation of these correlated flux measurements is used as an uncertainty estimate, typically $0.2$ Jy at all baselines. The uncertainties measured in this way broadly agree with the DRS-estimated values. The squared visibilities are finally calculated as $V^{2}(u,v,\lambda) = [F_{\rm targ}^{\rm cal}(u,v,\lambda) / F_{\rm targ}^{\rm cal}(u=0,v=0,\lambda)]^2$, where the ``zero-baseline'' flux is is the arithmetic mean of the photometric flux spectra measured by each of the 8.1m UTs. 

 The photometric flux was initially reduced via incoherent processing in the DRS (using \code{corrFlux=FALSE}). This mode extracts the photometric flux passing through the $2\lambda/D$ pinhole in each UT (0.61'' at 12 \micron). This is not computed for each observing block, as the $N$-band photometry cycle adds 10 minutes to each observation, but once per epoch we record the photometry. The $N$-band photometry we obtain from the DRS is a factor $\sim 3$ larger than expected from the MIDI and VISIR observations in T14, $36\pm 4$ Jy vs $12 \pm 1$ Jy at 12 \micron. We doubt temporal flux variations in the source, as none of the correlated fluxes at any spatial scale exhibit a similar change since 2008 (see \S\ref{sec:fluxvar}). 
When using EWS \citep[][]{jaffe2004b}, which was used previously for the MIDI observations, we extract a photometric flux of $12.4\pm 0.5$ Jy at 12\micron~for the same set of observations. 
This indicates that the photometric flux only exhibited a change due to the spatial filter used in each software; EWS employs a narrow Gaussian filter while DRS employs a wider top-hat. To compare consistently to the MIDI data, the EWS value is used.

We assume the calibration stars are symmetric and have zero closure phase on all phase triangles -- any deviations from zero represent instrumental phase errors.
 As a first step in closure phase calibration, deviations from zero phase in the calibrator $\delta \phi_{\star, ijk}(\lambda)$ are subtracted from the target phase: $\phi_{ijk,\rm targ}^{\rm cal} = \phi_{ijk,\rm targ}^{\rm raw} - \delta \phi_{\star, ijk}$. A typical MATISSE observation cycle includes 4 configurations of the BCD which serve to calibrate the closure phase. The varied BCD configurations (called out-out, in-in, in-out, out-in) should be identical save for sign flips on individual closure loops (as $\phi_{ijk} = -\phi_{ikj}$). 
We then average the star-calibrated closure phases. 
 We first calculate the temporal mean value for each individual BCD configuration, as they are each repeated a number $N_{\rm cycles}$ times. Finally the mean of the four BCD configurations serves as the closure phase value at each wavelength, and the standard deviation is used as an estimate of our closure phase uncertainty (on the order of $15^{\circ}$ for Circinus, on the order of $1^{\circ}$ for HD120404).  We note, however, that all closure loops which include the $\sim130$ m baseline, UT1-UT4, have systematically higher uncertainties due to the low signal-to-noise ratio (S/N) correlated flux on this baseline. The uncertainty is on the order of $50^{\circ}$ for Circinus, which means that only the closure triangles UT1-UT2-UT3 and UT2-UT3-UT4 provide high-precision phase information.

We have measurements in a total of 25 MATISSE exposure cycles, corresponding to 150 correlated flux measurements and 100 closure phase measurements. We define a position angle in the $uv$-plane as $\tan \psi = v/u$; we have sampled essentially all $\psi$ between $0$ and $110^{\circ}$, although the sampling is not uniform. This becomes especially noticeable on the longer baselines ($>100$ m). Two long-baseline regions at $\psi\approx[10,40]^{\circ}$ and $\psi\approx[80,110]^{\circ}$ are highly sampled, while a more sparse region is present between $\psi=45^{\circ}$ and $60^{\circ}$. On the shorter baselines, no such gaps are present.

\subsection{MIDI observations}
\label{sec:midi}
We include short-baseline MIDI observations from T14. These short baselines provide the small spatial frequencies necessary for imaging or modeling of the large-scale structure in Circinus. These data were reduced using the MIDI Expert Work Station \citep[EWS;][]{jaffe2004b}. The exact procedure is given in T14. These data contain the correlated flux, the visibility amplitude, and the wavelength differential phase. We calculate the squared visibility as $V^2 = (F_{\rm corr} / F_{tot} )^2$. Both the MATISSE and MIDI data have been calibrated with the same calibration star, HD120404. The MIDI data do not provide closure phases, so we select only a small number of AT baselines rather than fully incorporating the MIDI $uv$-coverage. We selected the baselines to have (\textit{i}) a projected baseline $< 35$m; and (\textit{ii}) $u,v$ spacing of at least 8.1~m (the UT-diameter). This leaves us with 18 baselines from the small configuration. In the MATISSE OIFITS format, these 18 baselines correspond to 12 closure phase loops, which we give as $0 \pm 180^{\circ}$ such that these nonexistent closure phases have no weight on imaging. This assumption is supported by the closure phase measurements of the VLT spectrometer and imager for the mid-infrared (VISIR) sparse-aperture-masking data.

\subsection{VISIR sparse-aperture-masking data}
\label{sec:visirsam}
Circinus was observed with the sparse-aperture-masking (SAM) mode of VISIR. The observations were taken in the $N-$band ($\lambda_0$ = 11.3 $\mu$m; Filter Name = 11\_3\_SAM) on 02 June 2017 (099.B-0484A). The data consisted on five observing blocks on the science with interwoven observations with the calibrator star HD\,125687. Each data set in the sequence SCI-CAL was observed with a DIT=142 milliseconds and NDIT=6 exposures. The data reduction consisted of two parts. The first one uses the ESOREX data reduction pipeline offered by ESO\footnote{\href{https://www.eso.org/sci/software/pipelines/visir/visir-pipe-recipes.html}{https://www.eso.org/sci/software/pipelines/visir/visir-pipe-recipes.html}}. It allowed us to correct for (i) the background, (ii) the bad pixels, (iii) to extract the interferograms from the chopping sequence and (iv) to center each frame on a 256$\times$256 pixel grid. Frames with low signal-to-noise or with bad cosmetics were discarded manually from the data. Once the interferograms were cleaned, we extracted the interferometric observables from them.

To obtain the squared visibilities and closure phases from the data, we used the CASSINI-SAMPip\footnote{\href{https://github.com/cosmosz5/CASSINI}{https://github.com/cosmosz5/CASSINI}} software \citep[see e.g.,][]{Sanchez-Bermudez_2020}. This algorithm fits the interferogram directly on the image plane, methods with similar performance based on fringe fitting are described by \citet{Greenbaum_2015} and \citet{Lacour_2011}. The code uses a Single Value Decomposition method to obtain the interferometric observables. The algorithm works with  monochromatic data and uses a sinc-filter for compensating the wavelength smearing of the broad-band VISIR filter. Each frame in the data was fitted independently. The uncertainties in the observables were obtained by averaging the observables of the six frames in each data set of science and calibrator, respectively. With the seven pin-holes mask available on VISIR, 21 squared visibilities and 35 closure phases were obtained per data set. The minimum baseline produced with the VISIR non-redundant mask has a length of 1.67 meters ($\lambda_0$/2B$_{\mathrm{min}}$ = 600 mas) and the maximum one a length of 6.28 meters ($\lambda_0$/2B$_{\mathrm{max}}$ = 184 mas), respectively. Figure \ref{fig:visir_uv} shows, as example, one snapshot of the recorded interferogram of the science target and the \uv-coverage obtained with our observations.  Once the raw observables were extracted, the data were calibrated by dividing the squared visibilites of the target over the ones of the calibrator star; the closure phases were calibrated by subtracting the closure phases of the calibrator from the ones of the target. 
Figure \ref{fig:visirsam} shows the calibrated observables versus spatial frequency. 

\begin{figure}[t!]
    \centering
    \includegraphics[width=0.55\textwidth]{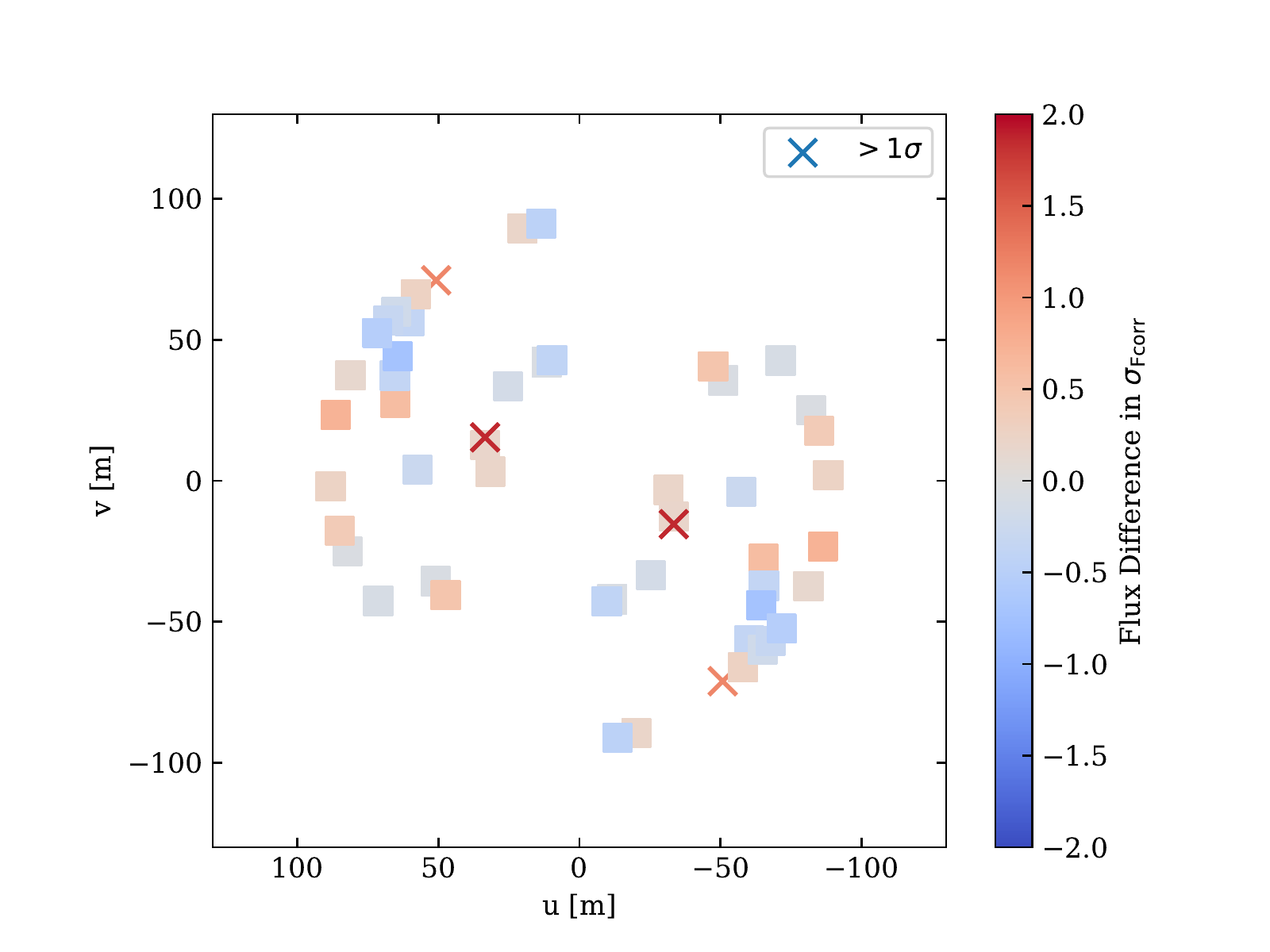}
    \caption{Comparison of MIDI and MATISSE correlated flux values on baselines cross-matched within 4 m. Color scale is difference in $\sigma_{\rm Fcorr}$ from the MATISSE observations. Only two $uv$-points are $1\sigma_{\rm Fcorr} < \Delta_{\rm Fcorr} < 2\sigma_{\rm Fcorr} $ discrepant between the MIDI and MATISSE observations spaced more than 10 years apart.}
    \label{fig:uv_fcorr_comparison}
\end{figure}

\subsection{Correlated flux stability}
\label{sec:fluxvar}
Combining the MIDI and MATISSE datasets taken $\geq$10 years apart depends on the assumption that both the structure and photometry of Circinus are stable in the same period. T14 reported possible flux variation of Circinus between 2008 and 2009. Moreover, there may be instrumental biases which are not properly calibrated. Therefore, we compare the correlated flux values taken using MATISSE in 2020 and 2021 with those at similar $u,v$ coordinates reported in T14. We identify and compare 30 baselines from MIDI and MATISSE which are within 4m in $u,v$ distance of each other; these are shown in Fig \ref{fig:uv_fcorr_comparison}. We find excellent agreement between the two epochs, with $>90\%$ of baselines consistent within the $1\sigma_{\rm Fcorr} \approx 0.2$ Jy calibrated correlated flux errors. Only two baselines are discrepant at 12~\micron~by $>1\sigma_{\rm Fcorr}$, but agree within $2\sigma_{\rm Fcorr}$. We find that there are no significant changes in correlated or total flux over the last $\geq 10$ years.

\subsection{Combination of MIDI and MATISSE data}  
\label{sec:combo}
Information at a large range of spatial frequencies is necessary for robust imaging of a source. The MATISSE UT observations have a shortest baseline of $\sim 30$ m, which causes structures larger than 82.5 mas to be resolved out at 12 \micron. Without MATISSE AT observations, we lack constraints on the large-scale structure. We know, however, that there is large-scale structure out to $\geq 600$ mas from MIDI, VISIR-SAM, and VISIR data \citep[T14, this work, and][respectively]{asmus2014}. In order to \textit{a)} avoid resolving out structure which is shown to be present, and \textit{b)} constrain the locations of small-scale structures, we perform the image reconstruction using a combination of MIDI and MATISSE data. The practice of including small spatial frequency data via modeling or data supplementation is common in imaging \citep[e.g.,][]{cotton2017}. The MIDI data do not include closure phase measurements, so as stated above we set the values to $0 \pm 180^{\circ}$ during imaging.  

We claim that such a data supplementation is valid in the case of Circinus for the following reasons. First, both the MIDI modeling and VISIR-SAM data show that closure phases are small on these scales ($\phi \leq 10^{\circ}$ in the T14 modeling; $\phi = 0.1 \pm 2.5^{\circ}$ in the VISIR-SAM data). Secondly, it is safe to combine the squared visibility measurements directly, as we show in \S\ref{sec:fluxvar} that the fluxes are stable on all scales over the last 17 years. Finally, we can combine the AT and UT data despite their different inherent spatial filtering because at 30 m baselines, the MIDI AT and MATISSE UT data give consistent correlated flux values, indicating that they probe the same structure. We finally note that the 18 included MIDI AT baselines represent only a small fraction of the imaging data, and serve primarily as a spatial constraint. The results of imaging both with and without the AT data are described further in Appendix \ref{sec:appImage}, but in summary the primary small-scale features remain stable in either approach.

\section{Image reconstruction}
\label{sec:imarec}
The primary advantage of MATISSE over MIDI is that the availability of closure phases makes it possible to reconstruct high-fidelity images. We employ the image reconstruction software, IRBis \citep[\textit{I}mage \textit{R}econstruction software using the \textit{Bis}pectrum;][]{hofmann2014,hofmann2016}, which was designed for MATISSE and is incorporated into the standard data reduction package. IRBis includes six regularization functions, two minimization engines, and myriad fine-tuning parameters such as the pixel scale, hyperparameter, and object mask. 
For the VISIR-SAM data, a completely independent image reconstruction process was carried out. We kept the image reconstruction for the MATISSE+MIDI data and that for the VISIR-SAM data separate due to dynamical range concerns, the different wavelength ranges, and because the spacial scales they measure are completely independent. 
We first focus on the image reconstruction of the MATISSE+MIDI data, with MIDI closure phases assumed to be $0\pm180^{\circ}$ (see \S\ref{sec:combo}). The image reconstruction for the VISIR-SAM data will be discussed in \S\ref{sec:samimarec}.

\subsection{MATISSE image reconstruction}
We select seven wavelength bins in which to produce independent images: $8.5 \pm 0.2$ \micron, $8.9 \pm 0.2$ \micron, $9.7 \pm 0.2$ \micron, $10.5 \pm 0.3$ \micron, $11.3 \pm 0.3$ \micron, $12.0 \pm 0.2$ \micron, and $12.7 \pm 0.2$ \micron. Any spectral information within each bin is averaged, producing a series of ``gray'' images. Each bin was imaged with a range of regularization functions and hyperparameters (hereafter $\mu$; essentially a scaling on the amount of regularization), with the best selected via a modified $\chi^2$ function: 
\begin{equation}
    q = \frac{\alpha}{N_{V^2}} \sum_{i=1}^{N_V} \frac{ (V^2_{\rm obs} - V^2_{\rm model, i})^2 }{\sigma_{\rm V^2, obs}^2 } + \frac{\beta}{N_{\phi}} \sum_{j=1}^{N_{\phi}} \frac{ (\phi_{\rm obs} - \phi_{\rm model, i})^2 }{\sigma_{\rm \phi, obs}^2 },
    \label{eq:qrec}
\end{equation}
with $\alpha$ and $\beta$ serving as weights on either squared visibilities or the closure phases. In IRBis, there are three ``cost functions'' which vary the relative weighting of the closure phases and squared visibilities during the image reconstruction process. For cost function 1, $\alpha = \beta = 1$; for cost function 2, $\alpha = 0$ and $\beta = 1$. Cost function 3 is more complex, using the $\chi^2$ coming from the sum of the bispectrum phasors and the squared visibilities \citep{hofmann2022}; in essence replacing the closure phases in the second term of Eq. \ref{eq:qrec} with the bispectrum. We employed cost function 1 for the quality assessment of best-fitting images.
\begin{table}[ht!]
    \centering
    \caption{Final image reconstruction parameters}
    \footnotesize
    \begin{tabular}{r|cccccc}
        $\lambda$  & Reg.$^a$ & $\mu^b$ & FOV$^c$  & Obj. Mask$^d$ & Cost$^e$& $\chi^2$$^{~g}$  \\
        $[\mu$m$]$  & Func.    &       & [mas] & [mas]       &  Func.  & [V$^2$,$\phi_{T3}$]  \\\hline \hline
        $8.5\pm0.2$   & 2 & 0.5& 600& 120&3&[2.6,3.8]\\
        $8.9\pm 0.2$  & 1 & 0.01 &600 &160 & 3&[1.7,3.4]\\
        $9.7\pm0.2$   & 2 & 0.18 &600 &120 &3&[0.5,0.9] \\
        $10.5\pm0.3$  & 5 & 0.08 & 500 & 120 & 1&[0.7,0.3] \\
        $11.3\pm0.3$  & 3 & 0.51 & 500 & 120 & 1&[2.3,1.4]  \\
        $12.0\pm0.2$  & 3 & 0.51 & 500 & 120 & 1&[6.4,1.6] \\
        $12.7\pm0.2$  & 5 & 0.30 & 600 & 140 & 1&[7.2,1.0] \\ 
    \end{tabular}
    \tablefoot{
    \textit{a}: the IRBis regularization function; \textit{b}: the weight on the regularization function (AKA the hyperparameter); \textit{c}: the field of view of the reconstructed image; \textit{d}: the radius of the object mask employed by IRBis in mas; \textit{e}: the cost function used in reconstruction, as described in Eq. \ref{eq:qrec} and in \citep{hofmann2022}; \textit{g}: the $\chi^2$ terms from the final images entering Eq. \ref{eq:qrec} for the squared visibilities and closure phases, respectively.}
    \label{tab:tab2}
\end{table}

\begin{figure*}
    \hspace{-2cm}
    \includegraphics[width=1.2\textwidth]{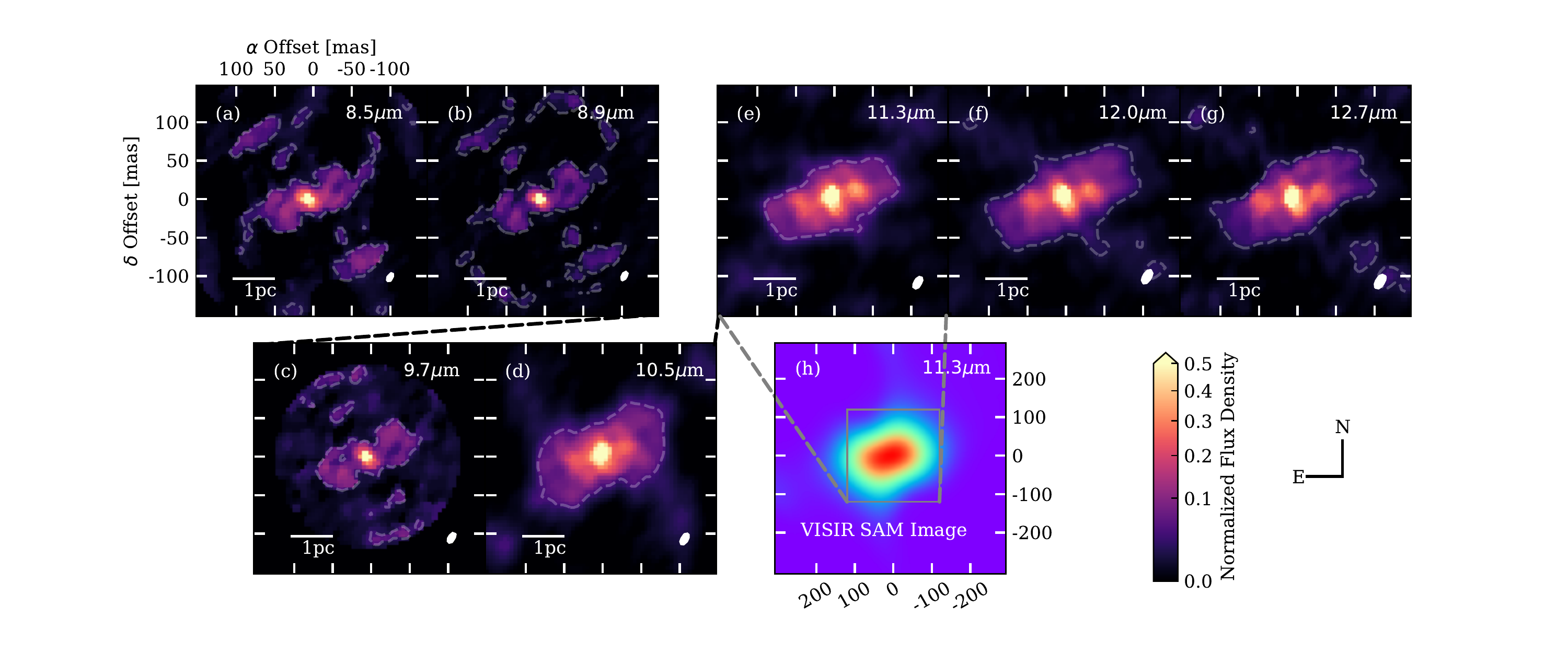}
    \caption{Compilation of MATISSE images reconstructed independently in each wavelength bin. Pixel scale and field of view are matched in all panels. The FWHM of the beam is shown in the bottom right corner of each panel. The top row consists of images of the continuum emission, and the bottom row holds the images within the Si absorption feature. The bottom row also includes the reconstructed VISIR-SAM image (panel \textit{h}), which has the field of view of the MATISSE images overlayed. 
    All images are scaled to the power of 0.6. Contours are drawn at $5\times$ the mean image error in each wavelength channel (see \S\ref{sec:imerrors}). }
    \label{fig:imgrid}
\end{figure*}

\begin{figure*}
    \centering
    \includegraphics[width=1\textwidth]{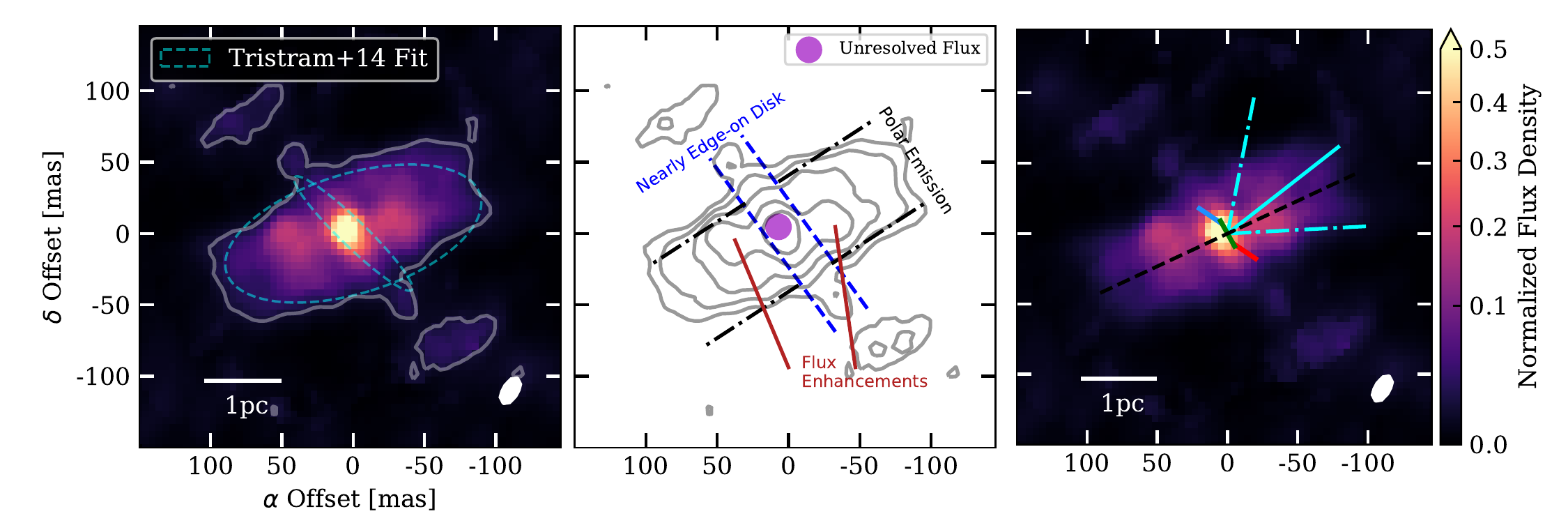}
    \caption{$N$-band continuum image and component labels. In the \textit{(left)} panel, we show the flux-weighted mean MATISSE image -- a proxy for an $N$-band continuum image. Contours are drawn at $5\times$ the flux-weighted mean of the individual image errors. The cyan dashed ellipses represent the FWHM of the Gaussians fitted to MIDI observations of Circinus by T14. 
    In the (\textit{center}) panel we show the same image as a contour map with levels at $[5,10,20,40,80] \times$ the $5\times$ the flux-weighted mean of the individual image errors. Key morphological features are labeled: the 1.9 pc disk with $i \gtrsim 83^{\circ}$, the polar emission, and the polar flux enhancement. 
    In the (\textit{right}) panel we show the same image with the \citet{greenhill2003} masers overplotted. 
    The black dashed line represents the direction of the radio jet \citep[][]{elmouttie1998}.
    The cyan lines show the central PA and opening angle of the optical ionization cone \citep[][]{fischer2013}. 
    Pixel scale and field of view are matched in both panels. All images and contours are scaled to the power of 0.65. The FWHM of the beam is shown in the bottom right corner of each panel. North is up and east is to the left. 
    }
    \label{fig:continuum}
\end{figure*}

In order to produce images, we performed a grid search of the IRBis parameters, varying the field of view (FOV), the pixel number, the object mask, the regularization function, the hyperparameter $\mu$, the cost function, and the reduction engine (ASA-CG or L-BFGS-B, see \citet{hofmann2016} for more details). We use uniform weighting in the $uv$-plane, corresponding to \code{weighting=0} in IRBis. An initial best image is selected in each wavelength bin using Eq. \ref{eq:qrec}, 
and a follow-up round of imaging using the best regularization function and pixel scale is performed. 
Regularization is a crucial component of ill-posed problems such as image reconstruction where the number of free parameters ($\approx N_{\rm px}^2$) is much larger than the number of data points. Regularization is the enforcement of an a priori constraint (e.g., smoothness, compactness, edginess, etc.) to prevent overfitting, but the strength of enforcement is set by the hyperparameter.
Starting from the initial images, we finely vary the hyperparameter to construct $L$-curves-- diagnostic comparisons between the amount of regularization and the residuals of the reconstruction \citep[first applied by][]{lawson1995}.
One identifies the ``elbow'' of the curve as the image with optimal regularization parameters. This selection is necessary to strike a balance between over-regularization and allowing too many image artifacts to manifest. We give the final parameters for the reconstructions in Table \ref{tab:tab2}. We note that different regularization functions in the same wavelength bin often result in very similar morphology, implying that the result is robust and simply not a consequence of regularized noise. Furthermore, the cost function has little effect on the final morphology or image quality and primarily aids convergence. 
We show the reconstructed images in Fig. \ref{fig:imgrid}, separating the continuum images from those inside the Si absorption feature. We also show the flux-weighted mean of the continuum images in Fig. \ref{fig:continuum} which represents an $N$-band image. 
Finally, we show the fit quality of each image in Figs. \ref{fig:cflux0} - \ref{fig:t3phi3}; we simulate the correlated fluxes and closure phases represented by each image at each \uv-point and compare to the observed data. We see that overall the images trace the closure phase and correlated flux spectra well, although specific wavelengths at a handful of \uv-coordinates are discrepant.

\subsubsection{Image error analysis}
\label{sec:imerrors}
We use the values in Table \ref{tab:tab2}, which represent the ``best" reconstruction parameters, to estimate the image-plane uncertainties. 
We do this through delete-$d$ jackknife resampling of our \uv-coverage \citep[the method is developed in][]{shao1986}.
In each Monte Carlo realization we randomly discard 10\% of each the squared visibilities and the closure phases (i.e., 15 squared visibilities and 10 closure phases). This choice satisfies the criterion for being asymptotically unbiased: $\sqrt{n} < d < n$, where $n$ is the sample size and $d$ is the number of deleted elements. We then perform the image reconstruction at each wavelength using the parameters given in Table \ref{tab:tab2} and save the results. After 100 realizations, we calculate the median and standard deviation in each pixel of each image. The median image at each wavelength is used as the final image shown in Fig \ref{fig:imgrid}. 
The standard deviation image serves as an error map with which we calculate the S/N at each pixel. The error maps and S/N maps are given in Appendix \ref{sec:appErrors}. We use the median image at each wavelength for our morphological analysis. The patchiness of the extended structure at 12.0 and 12.7 \micron~is moreover confirmed through measurement of flux within a 14 mas aperture at several points in the polar emission. Taking the image errors into account, the differences between adjacent bright and dark regions (e.g., at [(51.6, 23.4), (51.6, 46.9)] mas and [(18.8, 37.5), (32.8, 37.5)] mas from the image center) are $\geq 2\sigma$.

\subsubsection{Morphology}
In the final, independently reconstructed images (shown in Fig. \ref{fig:imgrid}), we find several consistent and key features. We discuss each below and have labeled them in Fig \ref{fig:continuum} for reference.

    \textbf{A central disk-like component.} This component is resolved in the NE-SW direction ($\approx1.9$ pc across), but unresolved at all wavelengths in the NW-SE direction. Its orientation varies slightly in the different wavelength channels: along PA$_{disk} \approx 45^{\circ}$ in the 8.5 and 8.9 \micron~images; and along PA$_{disk} \approx 30^{\circ}$ in the images red-ward of the Si feature.
    
    \textbf{Central, unresolved flux.} It is $\approx 10$ mas ($= 0.2$ pc) to the NE of the center of the disk in the 12 \micron~image. This is the brightest feature of the image at all wavelengths. 
    
    \textbf{Significant extended emission in the polar direction} (PA $\sim 295^{\circ}$), perpendicular to the maser emission and roughly aligned with the radio jet (see Fig. \ref{fig:continuum} for the orientations). The large-scale emission is more prominent at longer wavelengths. In the $11.3, 12.0$, and 12.7 \micron~images, the extended emission is approximately symmetric about the photo-center, and it is roughly 4 $\times$ 1.5 pc. This emission is notably not smooth, and shows patchiness far above the noise level. 
    
    \textbf{Two bright components, forming a rough line with the photo-center} at PA$_{E-W}=-80^{\circ}$ and superimposed on the polar emission, are observed for the first time. These substructures become more prominent at longer wavelengths, but are nonetheless present in all channels. They each extend to $\sim 65$ mas ($=1.2$ pc) from the center and are both roughly 30 mas across at 12 \micron.

At each wavelength we measure the flux contributions of the unresolved component, the disk, and the polar emission. These values are the total flux inside elliptical apertures placed at the image center with dimensions (10 $\times$ 10 mas), (100 $\times$ 10 mas) with PA$=45^{\circ}$, and (220 $\times$ 120 mas) with PA=$-65^{\circ}$, respectively. The contributions from the disk and unresolved component are subtracted from the largest aperture. Similarly, the contribution from the unresolved component is subtracted from the disk aperture. These values are presented in Table \ref{tab:fluxComponents}. The fractional contribution of the unresolved component increases to shorter wavelengths, indicating that it contains relatively hot dust; conversely, the polar emission contribution increases at longer wavelengths because it is cooler.

\begin{table*}[]
    \centering
    \small
    \caption{Measured fluxes of circumnuclear dust components}
    \begin{tabular}{c|rrr|rrr}
$\lambda$ & F$_{\rm polar}$ & F$_{\rm disk}$ & F$_{\rm unres.}$ & $f_{\rm polar}$ & $f_{\rm disk}$ & $f_{\rm unres.}$ \\
$[\mu$m$]$ & [Jy] & [Jy] & [Jy] &[\%] &[\%] &[\%] \\ \hline\hline
$8.5 \pm 0.2$ &$1.84 \pm 0.07$ & $0.46 \pm 0.03$& $0.39 \pm 0.01$ &
$42.5 \pm 1.5$ & $10.5 \pm 0.6$& $9.0 \pm 0.2$ \\
$8.9 \pm 0.2$ &$1.18 \pm 0.08$ & $0.29 \pm 0.04$& $0.36 \pm 0.01$ &
$37.4 \pm 2.5$ & $9.2 \pm 1.1$& $11.3 \pm 0.4$ \\
$9.7 \pm 0.2$ &$1.40 \pm 0.16$ & $0.28 \pm 0.06$& $0.28 \pm 0.02$ &
$48.6 \pm 5.6$ & $9.8 \pm 2.1$& $9.6 \pm 0.7$ \\
$10.5 \pm 0.3$ &$3.28 \pm 0.34$ & $0.60 \pm 0.09$& $0.29 \pm 0.02$ &
$64.7 \pm 6.6$ & $11.8 \pm 1.7$& $5.6 \pm 0.4$ \\
$11.3 \pm 0.3$ &$5.42 \pm 0.42$ & $1.00 \pm 0.11$& $0.49 \pm 0.03$ &
$63.9 \pm 4.9$ & $11.8 \pm 1.3$& $5.7 \pm 0.3$ \\
$12.0 \pm 0.2$ &$7.79 \pm 0.58$ & $1.61 \pm 0.17$& $0.77 \pm 0.04$ &
$64.0 \pm 4.8$ & $13.2 \pm 1.4$& $6.3 \pm 0.3$ \\
$12.7 \pm 0.2$ &$9.44 \pm 0.93$ & $2.04 \pm 0.28$& $0.96 \pm 0.07$ &
$61.5 \pm 6.1$ & $13.3 \pm 1.8$& $6.3 \pm 0.4$ \\
    \end{tabular}
    \tablefoot{Aperture fluxes (\textit{left}) and fractions of the total photometric flux (\textit{right}) for the polar emission, disk, and unresolved component in each image reconstruction. Fractional values do not sum to $100\%$ because some of the total flux is resolved out by the $\geq 30$ m baselines.}
    \label{tab:fluxComponents}
\end{table*}

There are several features which, while containing a large amount of flux, we consider to be artifacts of the image reconstruction process. A first, simple criterion is to take a S/N cut of $\sigma_{\rm image} \geq 3$, using the errors derived in \S\ref{sec:imerrors}. 
This simple cut agrees well with the following, more detailed considerations. 
If structures increase their size or radial distance from the photo-center linearly with wavelength, it is likely that they are artifacts of the \uv-coverage. This is complicated, however, as different wavelengths probe different temperatures in the thermal infrared, and thus real structures may become ``larger'' at longer wavelengths where cooler dust is observed. An example of an artifact which varies with wavelength is the pair of arc-like emission features $\sim100$ mas to the NE and SW of the photocenter in the 9.7 \micron~image. These appear to correspond to the secondary peaks of the dirty beam (\ref{fig:dirtyBeam}).
Finally, we assume that structures in the continuum should vary smoothly between adjacent imaging bins, and so we only consider those structures which are present in both the 8.5 and 8.9 \micron~images and those structures which are in all of the 11.3, 12.0, and 12.7 \micron~images to be real. We take the flux-weighted average of these five continuum images to produce a proxy for an $N$-band image. This is shown in Fig. \ref{fig:continuum}. The continuum-average image emphasizes consistent features of the images while suppressing artifacts.

\subsubsection{Effects of \uv-coverage on image morphology}
\label{sec:uvTest}
In this section we check what effects the attainable \uv-coverage could have on our final images. On the two longest baselines, we have no \uv-coverage for $\psi > 110^{\circ}$, and even more notably, we have no \uv-measurements on any baseline for $135^{\circ} < \psi < 180^{\circ}$. These \uv-holes are currently unavoidable due to VLTI delay line shadowing on the UT1-UT4 and UT2-UT3 baselines for Circinus at DEC =$-65:20:21$. While MATISSE can be used in a two-baseline configuration, we would not be able to measure the closure phases necessary for imaging. This \uv-region has been shown by T14 to be important nonetheless, as the disk-like structure present in their modeling is primarily constrained by long baselines in this direction. T14 reports MIDI measurements of the UT1-UT3 baseline ($\sim90$ m) in the \uv-region we cannot currently measure. In order to test the effects of including measurements at these $\psi$, we performed a second round of imaging, incorporating MIDI baselines with BL $\in [30,100]$ m, $\psi_{\rm MIDI} \in [100,180]^{\circ}$, and which are separated by at least 4 m in \uv-coordinates. These criteria resulted in 18 additional baselines with correlated flux measurements from MIDI. As there were no closure phases for these baselines, we use the same procedure as when including the MIDI AT measurements, setting $\phi_{\rm T3, MIDI} =  0 \pm 180^{\circ}$.

Adding these 18 MIDI UT correlated fluxes to the 150 MATISSE measurements and the 18 MIDI AT measurements, we produced independent images at 8.9 and 12.0 \micron. At 8.9 \micron~the resulting image is essentially unchanged, indiscernible by eye from the image shown in Fig. \ref{fig:imgrid}; an explicit comparison is shown in Fig. \ref{fig:app_utonly}. 
At 12.0 \micron, however, the disk becomes more prominent and changes position angle slightly, while all other features remain constant. 
 The disk-like structure in the initial imaging lies along PA$_{\rm disk} \sim 35^{\circ}$, while after the addition of MIDI UT baselines the same disk-like structure lies along $\sim 40^{\circ}$. The latter value more closely resembles the $46 \pm 3^{\circ}$ given in T14. However, given the size of the beam at 12 \micron, 9 mas, the disk orientation could quite easily vary in the image by $\sim3^{\circ}$. 
The overall differences in the image plane are small when we include these baselines, so we proceed in our analysis without the MIDI UT measurements. 
It is clear from T14, however, that these baselines are important to understand the size and orientation of the disk-like structure in Circinus, and the planned doubling of the VLTI delay lines will make closure phase measurements including these baselines possible.

\subsection{VISIR-SAM image reconstruction}
\label{sec:samimarec}
The VISIR-SAM data allowed us to reconstruct an image of the target's large scale at 11.3 $\mu$m.
We reconstruct the VISIR-SAM image and the MATISSE images separately, rather than combining the \uv-coverage for two reasons: first, the longest SAM baselines ($\approx 7$~m) are shorter than the shortest MIDI-AT baselines, meaning there is no overlap in measured visibilities; second, the SAM data exhibit squared visibilities $> 0.4$, which are much larger than the MATISSE values on longer baselines and would result in dynamical range issues during imaging. The data are nonetheless useful as a way to contextualize the MATISSE images and to measure the true extent of the polar emission.
We used BSMEM \citep{Buscher_1994, Lawson_2004} to reconstruct the VISIR-SAM image. This code uses a regularized minimization algorithm to recover an image from infrared interferometric data. The regularized optimization engine uses a trust-region gradient-descent method with entropy (i.e., the sum of the logarithm of the pixel values in the image grid) as regularization function. Images were reconstructed using squared visibilities and closure phases simultaneously. The reconstructed image used a pixel scale of 15 mas in a pixel grid of 512 x 512 pixels. The code converges to a $\chi^2$ close to unity. Figure \ref{fig:imgrid} shows our VISIR-SAM image.

\begin{figure*}
    \centering
    \includegraphics[width=0.45\textwidth]{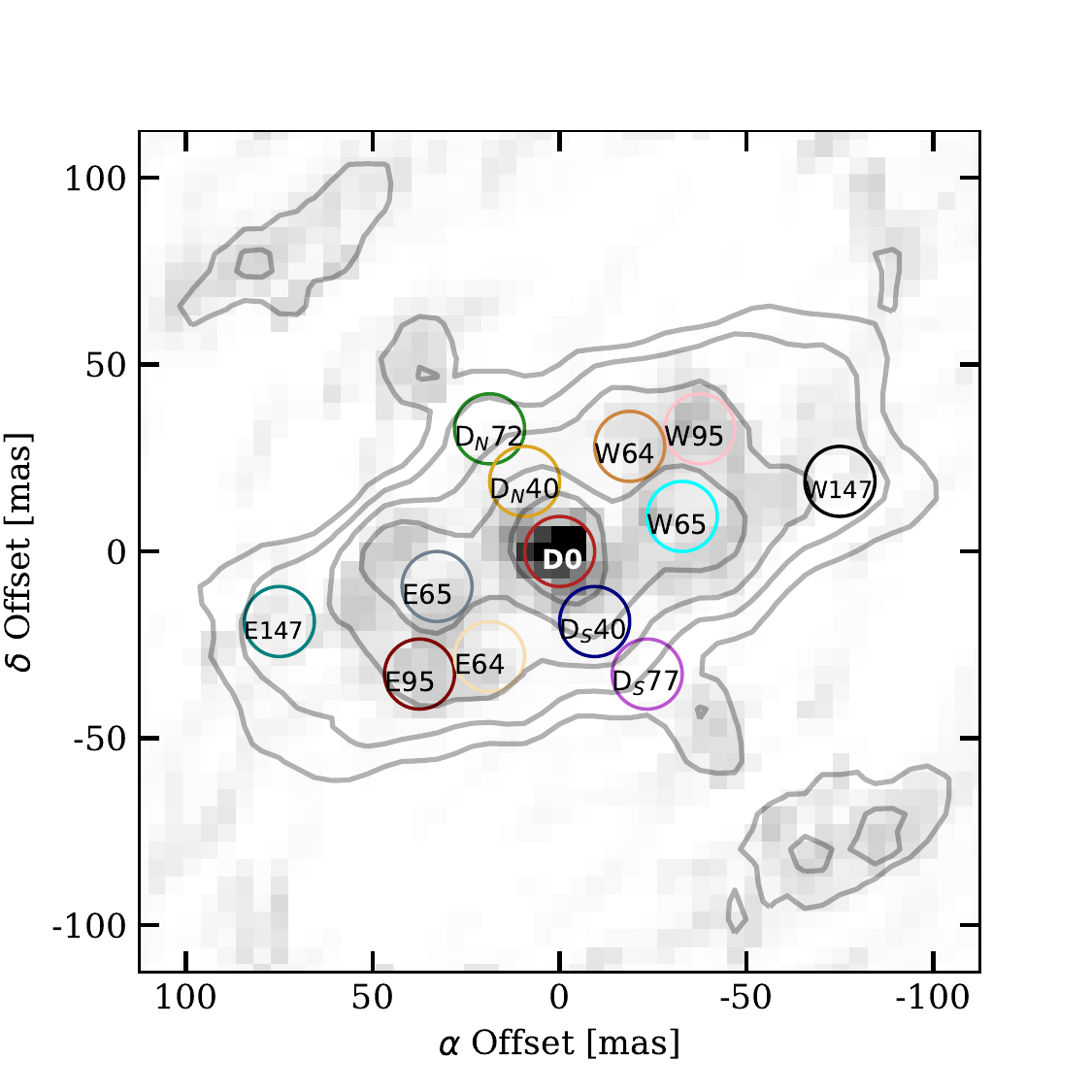}
    \includegraphics[width=0.5\textwidth]{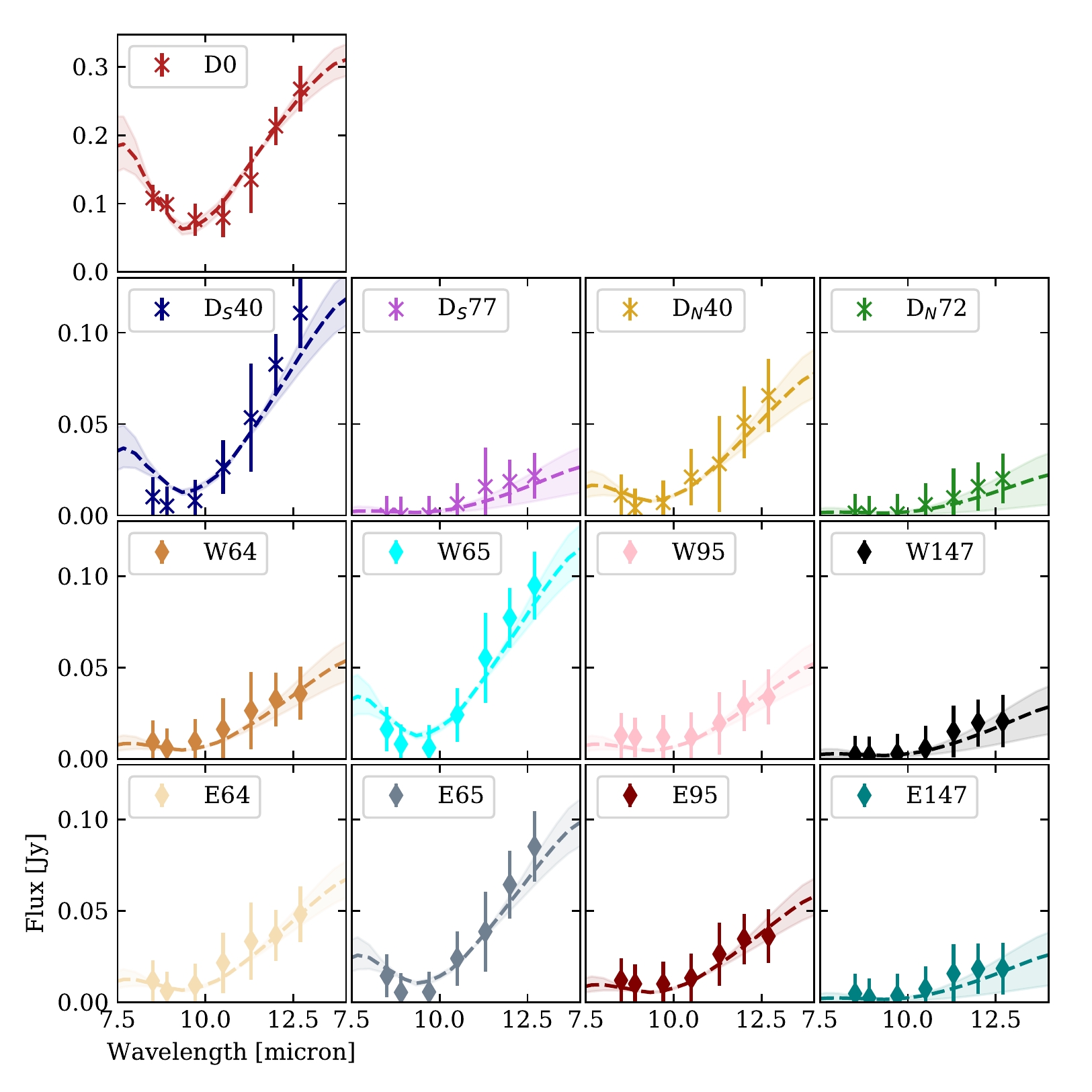}

    \caption{In the (\textit{left}) panel we show the continuum image contours on top of the 8.9~\micron~image. Overplotted on these are the 13 apertures we used for SED extraction. 
    The contours are at $[5,10,20,40,80]$ times the continuum image error (see Fig. \ref{fig:continuum}). All images and contours are scaled to the power of 0.65. In the (\textit{right}) panel we show the extracted mean flux at each wavelength in each aperture as well as the best-fitting blackbody functions to each SED. These fits had uniform priors on $T$ and Gaussian priors on A$_{\rm V}$ with $\mu = 28.5$ mag and $\sigma=8.1$ mag. The shaded regions represent $1\sigma$ uncertainties as estimated from the posterior probability distributions.}
    \label{fig:sed}
\end{figure*}

\section{Measuring the dust temperature distribution}
\label{sec:temperature}
The images produced above supply not only morphological information, but also information about the temperature and optical depth of the dust in different regions. In this section, we fit one-temperature blackbody models with extinction to a series of apertures. As shown in \citet{gamezrosas2022}, Gaussian modeling, point-source fitting, and image reconstruction all resulted in similar extracted SEDs in NGC 1068; therefore we can with confidence use the extracted apertures from our reconstructed image to undertake a temperature analysis of Circinus. We first convolve the images to the beam of the lowest resolution reconstruction (12.7 \micron, corresponding to 10.1 mas). 
The individual images are aligned using cross-correlation before SED extraction, but effectively the photo-centers are simply matched.
Then we define 13 apertures (shown in Fig. \ref{fig:sed}) which are 5 px (23.4 mas) in diameter and which do not overlap; their exact locations were chosen by hand to cover key features of the disk and polar emission.
This is $> 2\times$ larger than the lowest resolution ``beam'' in our images, and in this way we do not make claims based on any hyper-resolved features.
 We extract the mean flux from each aperture in each image and estimate the flux error from the same apertures on the error maps estimated in \S\ref{sec:imerrors}.
Finally, we add the calibration error of the total photometric flux at each wavelength in quadrature to the extracted flux error.

We fit a single blackbody (BB) curve with absorption to each aperture-extracted spectrum with the form\footnote{We do not include a ``graybody'' emissivity here because the two-parameter model provided robust fits to the spectra.}
\begin{equation}
    I(\lambda, T, A_v) =  BB_{\nu}(\lambda,T) e^{ -A_v / 1.09  \tau(\lambda)/\tau_v }  ,
    \label{eq:bb}
\end{equation}
where $\tau(\lambda) / \tau_v =\kappa(\lambda) / \kappa_v $, and we use the standard interstellar medium $\kappa(\lambda)$ profile from \citet{schartmann2005} which is based on the standard interstellar medium profile of \citet{mathis1977}. 
Fitting of $T$ and A$_{\rm V}$ was done in two iterations using Markov Chain Monte Carlo sampling with the package \pack{emcee} \citep[][]{foreman-mackey2013}. Final values in each iteration are the median of the marginalized posterior probability distribution. The 16th and 84th percentiles of the resulting temperature and extinction distributions are used as the $1\sigma$ fit uncertainties\footnote{Valid only because the resulting distributions are approximately Gaussian.}.

In the first iteration, we use uniform prior probability distributions with $T \in [100,600]$ K and A$_{\rm V} \in [0, 100]$ mag. We find that the extinction does not vary significantly across the field. The central aperture, fit with the smallest uncertainties, shows A$_{\rm V} = 28.5_{-7.7}^{+8.5}$ mag which is $\tau_{9.7} = 2.0_{-0.5}^{+0.6}$ using the mass-extinction profile from \citet{schartmann2005}. The other apertures have nominally higher values, but large uncertainties which make the differences insignificant. Only D$_S 40$, W65, and E65 show differences $> 1\sigma$ from the central value.

In the second iteration, we use again a uniform prior for temperature ($T \in [100,600]$ K), but a Gaussian prior for A$_{\rm V}$ with $\mu = 28.5$ mag and $\sigma=8.1$ mag based on the fit to D0 in the first iteration. The central aperture, D0, serves as a good estimate of the overall extinction because it \textit{a)} has the highest S/N and \textit{b)} has significant flux on both sides of the Si absorption feature. The resulting temperatures are consistent with the unconstrained case but are typically lower. The qualitative behavior of the temperature distribution is unchanged, but the fitted uncertainties are greatly diminished due to the degenerate nature of A$_{\rm V}$ and T for a fit to the Rayleigh-Jeans tail of the Planck function. In the following discussion, we therefore use the values from the second iteration. We show the best fit parameters for each aperture in each iteration in Table \ref{tab:sed_params}.

We do not find evidence of an extinction gradient across the field, indicating that there is a relatively uniform screen of foreground absorption. In the first fitting iteration, with A$_{\rm V}$ allowed to vary, the mean extinction values are similar to the east and west. In the second iteration, we restricted A$_{\rm V}$ around 28.5 mag, to get better constraints on the temperature. Based on Hubble $K$-band imaging, \citet{wilson2000} estimated an extinction of A$_{\rm V} = 28\pm7$ toward a compact ($<2$ pc) nucleus. \citet{burtscher2016} measured a value of A$_{\rm V} = 27.2\pm3$ using SINFONI in the $K$-band. \citet{roche2006} found $2.2\leq \tau_{9.7} \leq 3.5$ using T-ReCS on Gemini-South. Previous measurements are nearly identical to the fitted value in D0, $28.7_{-7.7}^{+8.5}$ mag, and furthermore consistent with the rest of the field. Uniform absorption, however, is in contrast to the $\Delta \tau = 27$~arcsec$^{-1}$ gradient across the polar emission measured by T14. This discrepancy is puzzling, but we recognize there are major differences between our approach and that of T14. Specifically, T14 used differential phases and Gaussian modeling due to the lack of closure phase data. Their differential phases were measured on the UT and AT baselines, and thus probe larger-scale material than the MATISSE UT closure phases alone. On the other hand, we use no differential phases and had to assume an unconstrained AT closure phase value of $0\pm180^{\circ}$. 
However, we note that on the UT baselines (probing $\lesssim 1$~pc scales), the 9.7 \micron~closure phases are well matched by our images \textit{without} an extinction gradient. The phase signals are instead produced by small-scale structure that was smoothed out in the Gaussian modeling approach of T14. The two approaches emphasize different aspects of the data, but differential phases could be included with the closure phases in future work in chromatic image reconstructions. Future closure phase measurements at 9.7 \micron~are required on shorter baselines (e.g., with MATISSE AT observations) to directly measure the Si absorption across the large-scale component.

\begin{figure}
    \centering
    \includegraphics[width=0.55\textwidth]{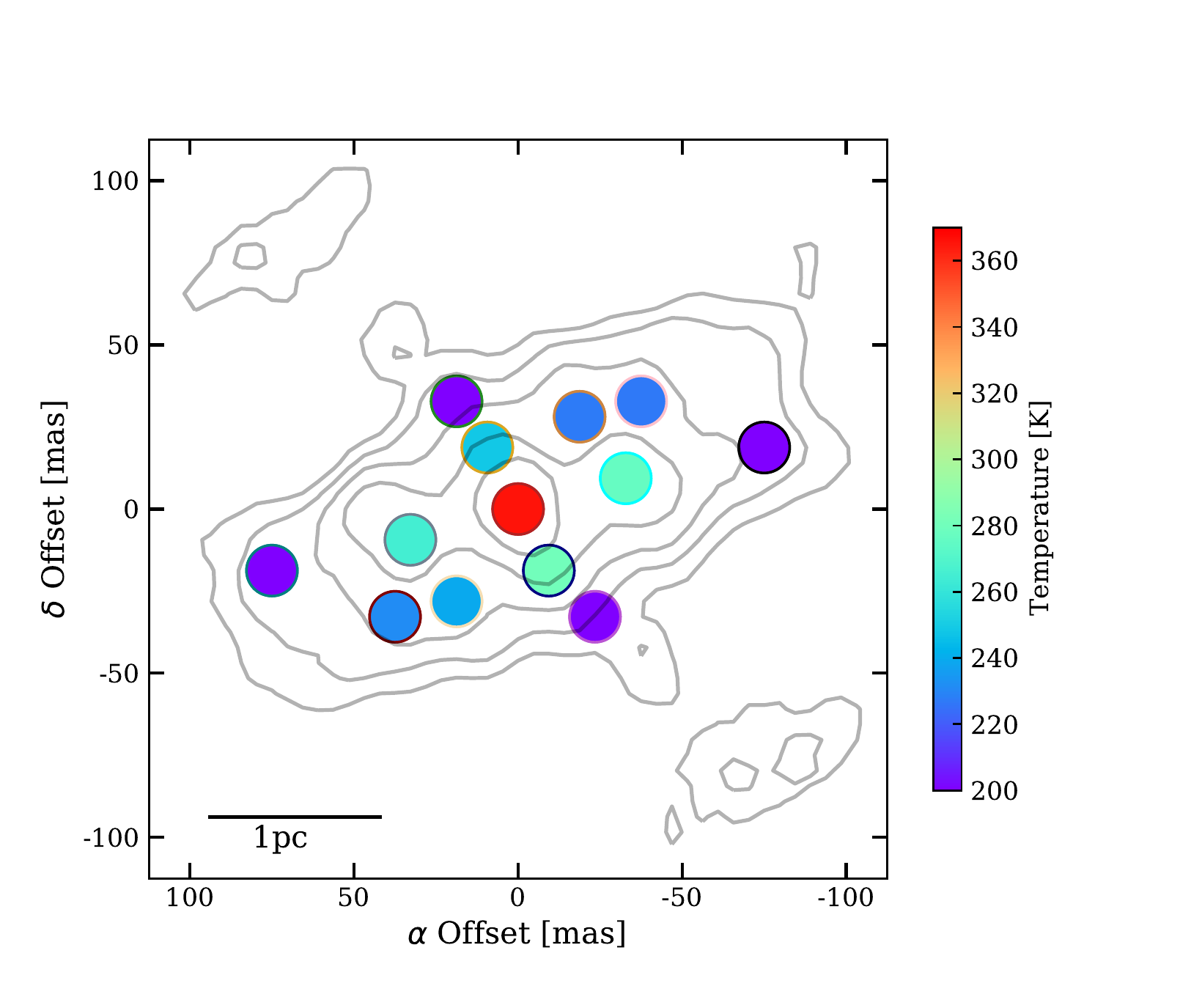}
    \caption{Two-dimensional temperature distribution as fitted in each of 13 apertures. Colors at the edge of each circle match those given by the aperture definitions in Fig. \ref{fig:sed}. Temperatures within the polar extension remain high even at large distances from the center when compared to the ``disk'' apertures. The contours are at $[5,10,20,40,80]$ times the continuum image error and scaled to the power of 0.65 (see Fig. \ref{fig:continuum}).
    }
    \label{fig:temperature_map}
\end{figure}

We separate the apertures into two rough categories based on their locations. Those oriented NE and SW from the photocenter at PA$\sim 30^{\circ}$ are labeled as ``disk'' apertures, based on the presence of a thin disk-like structure in both our reconstructed images and in the Gaussian modeling of T14. The other points, extending NW and SE from the photocenter are labeled ``outflow'' apertures, as they lie in the direction of the polar extension. The extracted spectra and the fitted blackbody curves (with uncertainty estimates as shaded regions) are shown in Fig. \ref{fig:sed} .  The two-dimenstional temperature distribution based on the fits is shown in Fig. \ref{fig:temperature_map}. 
We find that on average the ``disk'' apertures show a much steeper temperature falloff with projected distance than the ``outflow'' apertures.

\subsection{Temperature gradient analysis}

In current modeling, the dust in the outflow is anisotropically illuminated by a face-on accretion disk. We can use the temperature profile of the outflow to characterize the dust environment. 

We begin with a comparison to the simple analytic model of \citet{barvainis1987}:
\begin{equation}
    T_{gr}(r) = 1650 \Big(\frac{L_{\rm acc}}{r^2} \frac{{\rm pc}^2}{10^{10}{\rm L}_{\odot}} \Big)^{1/5.6} e^{-\tau_{\rm uv}/5.6} {\rm K},
    \label{eq:barvainis}
\end{equation}
where $L_{acc}$ is the luminosity of the accretion disk in $L_{\odot}$, $r$ is the distance from the accretion disk in parsec, and $\tau_{\rm uv}$ is the optical depth to the ultraviolet continuum. Here we use $L_{\rm acc} = 6 \times 10^9 L_{\odot}$, which is the lower bound on estimates of the accretion disk luminosity in Circinus ($6\times10^9 - 7 \times10^{10} L_{\odot}$), as inferred from X-ray \citep{arevalo2014, ricci2015}, IR (T14) and optical \citep{oliva1999} observations. We also use $A_{V} = 40 {\rm~mag}\rightarrow \tau_{\rm uv} = 7.2$, set roughly by the mean of the first-iteration fitted extinction values in Table \ref{tab:sed_params} and converted using the dust extinction curve of \citet{schartmann2005}, but we note that the best-fitting value, A$_{\rm V} = 28.5_{-7.7}^{+8.5}$ mag, would result in even higher temperatures at a given radial distance. 
With these assumed values, we compare the radial temperature profile of the optically thin, continuous dust environment described by \citet{barvainis1987} to the fitted SED temperatures of the ``outflow'' in Fig. \ref{fig:profiles}. At all radii, the Eq. \ref{eq:barvainis} temperatures are larger than the measured Circinus temperatures by a factor of $\sim 2$. This is not completely unexpected, as the inefficient re-radiation by the dust grains in the \citet{barvainis1987} model leads to higher temperatures at large radii; this model should be considered an upper limit on the dust temperatures at a given radius. Moreover, the \citet{barvainis1987} model does not take the anisotropy of the radiation into account. We also plot the expected temperature profile arising due to the simplest case of radiation equilibrium for perfectly efficient blackbodies as given in \citet{tristram2007} for comparison.

\begin{table}[ht!]
    \centering
    \small
    \caption{Fitted blackbody parameters for each of the 13 image-extracted spectra}
    \begin{tabular}{lllll||ll}
&&&First & Iteration & Second & Iteration \\ \hline
Aperture & Dist. & Temp. & A$_{\rm V}$ & $\tau_{9.7}$ & Temp. & A$_{\rm V}$\\ 
& [pc] & [K] & [mag] & &[K]&[mag]\\
\hline \textbf{Disk} &&&&\\\hline 
D0 & 0.00 & $\mathbf{367_{-26}^{+30}}$ & $\mathbf{28.5_{-7.7}^{+8.5}}$ & $2.0_{-0.5}^{+0.6}$ & -- & --\\
D$_{S}$40 & 0.40 & $358_{-56}^{+68}$ & $66.5_{-22.8}^{+22.3}$ & $4.6_{-1.6}^{+1.5}$ &$281_{-17}^{+17}$ &$33.3_{-7.1}^{+7.0}$ \\
D$_{N}$40 & 0.40 & $297_{-50}^{+64}$ & $58.2_{-29.0}^{+28.9}$ & $4.0_{-2.0}^{+2.0}$&$249_{-17}^{+17}$ & $29.8_{-7.7}^{+7.7}$\\
D$_{S}$77 & 0.77 & $221_{-43}^{+49}$ & $60.2_{-36.7}^{+28.1}$ & $4.1_{-2.5}^{+1.9}$&$198_{-29}^{+19}$ & $29.2_{-8.2}^{+8.1}$\\
D$_{N}$72 & 0.72 & $208_{-49}^{+51}$ & $58.8_{-35.7}^{+29.0}$ & $4.0_{-2.5}^{+2.0}$&$191_{-41}^{+22}$ & $29.2_{-8.0}^{+7.9}$ \\
\hline\textbf{NW}& \textbf{Polar} &\textbf{Ext.} &&\\\hline 
W64 & 0.64 & $246_{-35}^{+57}$ & $44.9_{-28.5}^{+36.1}$ & $3.1_{-2.0}^{+2.5}$&$228_{-16}^{+16}$ & $28.4_{-7.6}^{+7.9}$\\
W65 & 0.65 & $333_{-47}^{+64}$ & $58.6_{-21.1}^{+23.6}$ & $4.0_{-1.5}^{+1.6}$&$277_{-16}^{+17}$ & $32.0_{-7.0}^{+7.4}$\\
W95 & 0.95 & $244_{-34}^{+55}$ & $45.5_{-29.2}^{+34.4}$ & $3.1_{-2.0}^{+2.4}$&$227_{-16}^{+17}$ & $28.7_{-8.1}^{+8.0}$\\
W147 & 1.47 & $220_{-43}^{+50}$ & $58.2_{-35.2}^{+29.1}$ & $4.0_{-2.4}^{+2.0}$&$200_{-29}^{+19}$ & $29.0_{-8.1}^{+7.9}$\\

\hline\textbf{SE}& \textbf{Polar} &\textbf{Ext.} &&\\\hline
E64 & 0.64 & $261_{-38}^{+59}$ & $44.4_{-26.9}^{+34.2}$ & $3.1_{-1.8}^{+2.4}$&$240_{-14}^{+15}$ & $28.6_{-7.5}^{+8.2}$\\
E65 & 0.65 & $325_{-52}^{+66}$ & $61.6_{-25.5}^{+24.8}$ & $4.2_{-1.8}^{+1.7}$&$266_{-16}^{+17}$ & $31.4_{-7.3}^{+7.5}$\\
E95 & 0.95 & $248_{-33}^{+54}$ & $42.8_{-27.0}^{+34.8}$ & $2.9_{-1.9}^{+2.4}$&$231_{-11}^{+15}$ & $28.1_{-8.2}^{+8.1}$\\
E147 & 1.47 & $214_{-44}^{+50}$ & $54.9_{-35.0}^{+32.3}$ & $3.8_{-2.4}^{+2.2}$&$197_{-36}^{+21}$ & $29.0_{-8.0}^{+8.0}$\\

    \end{tabular}
    
    \tablefoot{
    $\tau_{9.7}$ is the simple conversion from A$_{\rm V}$ to the optical depth of the Si feature based on the $\kappa(\lambda)$ curve from \citet{schartmann2005} and is included only for comparison to previous results, namely T14. Projected distances in parsec are given from the central aperture, D0, with Circinus 4.2 Mpc away \citep{freeman1977}. We measure the inclination to be $i \gtrsim 83^{\circ}$, so the correction from projected to physical distance is small. The two rightmost columns are the results of re-fitting with a Gaussian prior on A$_{\rm V} = 28.5 \pm 8.1$, based on the initial fit to D0.
    }
    \label{tab:sed_params}
\end{table}

\subsection{Comparisons to radiative transfer models}
\subsubsection{Clumpy torus models}
Modern AGN ``torus'' modeling takes the clumpiness of the dust, as implied from infrared interferometry, as well as anisotropic illumination from the accretion disk into account. We compare the temperatures at different radial distances in the standard clumpy torus model of \citet{schartmann2008} to those fit in Circinus. These models consist of a wedge-shaped torus filled with randomly placed, optically-thick spheres of dust. The clump density falls off with radius, $r$, from the anisotropically illuminating source as $\rho \propto r^{-0.5}$ and the clump size increases as $a \propto r^{1.0}$. These models consist only of a puffy ``disk'' with half-opening angle $\theta = 45^{\circ}$, as they predate the observations of polar dust in Circinus. The clumpy torus models produce a range of dust temperatures as a function of radius which serve as a theoretical bound on the temperature distribution in the central few parsec.
The temperatures found by our blackbody fits are clearly within these theoretical bounds of the model, c.f.~Fig.~\ref{fig:profiles}. A similar result was already found by \citet{tristram2007, tristram2014}. 

\subsubsection{Disk + wind models}

More recently, \citet{stalevski2017,stalevski2019} undertook radiative transfer modeling of VISIR imaging data, the MIR SED, and MIDI interferometric data of Circinus. Their best-fitting model \citep[presented in][]{stalevski2019} consists of a compact, dusty disk and a hollow hyperbolic cone extending in the polar direction (hereafter disk+hyp).
In this modeling, a parameter grid for the radiative transfer models was searched such that the overall SED as well as the interferometric observables were well reproduced. This was not a model fit, but rather an exploration of the parameter space. 
For comparison with the MATISSE data, we started from the best model of \citet{stalevski2019} and varied its parameters with finer sampling of the parameter space. We significantly expanded the explored range of the parameters which define the clumpiness: the number of clumps (i.e. filling factor) and different random realizations of the clumps' positions (set by the "seeds" for the random number generator). Using the MATISSE \uv-coverage, we simulate the squared visibilities and closure phases of each model image and compute the $\chi^2$ to the data (the comparisons and resulting model are shown in Fig. \ref{appfig:models_observables}). This comparison placed constraints on the system inclination ($i \sim 85^{\circ}$), the hyperboloid opening angle ($\theta_{\rm OA} \sim 30^{\circ}$), the disk Si feature depth ($\tau_{\rm Si, DSK} \sim 15$), and the outer radius of the disk ($r_{\rm out} \sim 3$ pc). The closure phase comparison favored a small number of clumps. We then performed a finer parameter search based on these constraints, focusing on the filling factors of the disk and the hyperboloid. 
After comparing with the MATISSE data, the parameter values defining the boundaries of the model geometry remain unchanged (the dusty disk outer radius, angular width, optical depth; hyperboloid shell position, width and optical depth). However, our modeling converged on a smaller number of clumps ($30\%$ less than in the MIDI model) and found that random positions of the clumps have a significant impact on the quality of the fits. The selected model exhibits a sky covering fraction of 78\% due to the dust clumps at 0.55 \micron. We show in Fig. \ref{fig:profiles} the average dust temperature as a function radius; these are indicative temperatures obtained by averaging the local thermal equilibrium temperatures over the dust species and grain sizes.
We finally extract fluxes in each of the 13 apertures and fit blackbody temperatures using Eq. \ref{eq:bb} to the disk+hyp model grid at $\lambda \in [8.53, 8.91, 9.29, 9.70, 10.12, 10.56, 11.02, 11.50, 12.00, 12.52]$ \micron.

We compare the extracted model spectrum in each aperture to the observed spectra. We quantify this through the $\chi^2$ but do not perform any model fitting. These comparisons are shown in Fig. \ref{appfig:models_ff}. In the polar extension, the model and image extracted fluxes and temperatures agree well. The preferred model of \citet{stalevski2019} includes the polar dust flux-enhancements E-W of the center. Along the disk, and particularly in the central aperture, D0, we see significant discrepancies. 
The central aperture temperature is $\sim 100$ K less in the models than in the observations and the extracted flux is $\lesssim 10\%$ of the observations. These discrepancies may indicate that the model disk is perhaps too dense. The disk apertures D$_S77$ and D$_N77$ also show much lower observed temperatures than the model predicts, indicating that the model disk can be further improved. 
Given that the outer radius, angular width, average edge-on optical depth and inclination of the disk appear to be well constrained, it is likely that the disk is actually inhomogeneous, or perhaps with a gap, thus allowing more warm emission to escape. 
$LM$-band images at $\sim 3$ mas resolution are required to further constrain the disk component in modeling of Circinus. 
The MATISSE observations and imaging of Circinus agree very well with clumpy modeling, but it is beyond the scope of this work to place constraints on the specifics of a clumpy medium.

\section{Discussion}
\label{sec:discussion}

MIR interferometry of Circinus has revealed several major components of the thermal dust: a disk-like central emission, large-scale polar emission, and a central point source along the disk. Image reconstruction has recovered these features in unprecedented detail and brought forward new substructures. The morphological features are labeled in Fig. \ref{fig:sed}. In the following, we examine each of these features separately. After the discussion of the individual features, from the smallest scales to the largest, we discuss the overall morphology. 

The orientations of the central structures in Circinus are compared to those of the optical ionization cone and of the warped maser emission in the center. The well-studied optical ionization cone has a central axis along PA$_{\rm opt}=-52^{\circ}$ and a projected half-opening angle between $36^{\circ}$ and $41^{\circ}$ \citep[see, e.g.,][]{marconi1994,maiolino2000,wilson2000,fischer2013, mingozzi2019}. 
The observed ionized emission is not symmetric; it only extends toward the NW with no optical counterpart seen to the south, though a southern counterpart can be seen in the NIR \citep{prieto2004}. The ionization cone is thought to coincide with an outflow of dense material, driven by radiation pressure and fed by a gaseous nuclear bar \citep{maiolino2000, packham2005}. Notably, the O{[\sc III]} and H$\alpha$ emission in the ionization cone is much brighter along its southern edge ($PA \sim -90^{\circ}$). The ionization cone is observed out to $\sim 40$ pc from the nucleus \citep{wilson2000}.

The warped H$_2$O maser disk was separated by \citet{greenhill2003} into 3 components: the blueshifted emission ($0.11 < r \lesssim 0.4$ pc; PA$_{\rm maser, blue} = 56 \pm 6^{\circ}$), the central emission ($0 < r < 0.11 \pm 0.02$ pc; PA$_{\rm maser, central} = 29 \pm 3$ deg), and the redshifted emission ($0.11 < r \lesssim 0.4 $ pc; PA$_{\rm maser, red} = 56 \pm 6^{\circ}$). The central maser emission, which may trace the orientation of the accretion disk and the dense material around it, is nearly perpendicular to the radio jet axis \citep[PA$_{\rm jet} =$ 115 and 295 $\pm  5^{\circ}$;][]{elmouttie1998}
, which is not aligned with the central axis of the optical ionization cone. These orientation markers are shown in Fig. \ref{fig:continuum} for comparison to the MATISSE images.

\subsection{Unresolved central flux }
\label{sec:cps}

We find a central, bright component unresolved at all wavelengths ($\leq$6.7 mas at 8.5 \micron -- $\leq$10.1 mas at 12.7 \micron). It is in the same position relative to the image photocenter in each reconstruction, and is therefore likely the same unresolved object present throughout. This point source is consistently found $\approx 10$ mas to the NE of the photocenter of the disk. Our central aperture, D0 (Fig. \ref{fig:sed}, Table \ref{tab:sed_params}), is centered on this point source and the extracted fluxes are brighter than the surrounding features by more than a factor of 2. We find that the fitted blackbody is relatively hot, $366.7_{-26}^{+30}$ K.  While this source was well-fit by a single blackbody, we note that this is difficult to motivate physically and only serves as an estimate.

These results are similar to those of T14, who found a central unresolved component lying along the disk-like component. Their point source was shifted 14 mas to the NE of the disk-center, similar to the 10 mas which we find. T14 measured the temperature of this component to be $317\pm 22$ K, which is $\sim 2\sigma$ lower than our measured temperature. 
The temperature difference is perhaps a result of the overlapping contributions of the three Gaussian components in T14, while we fit an isolated mean temperature at each extraction location. Nonetheless, no directly visible hot ($\gtrsim 900$ K) dust is found by either T14 or this work.

The central aperture is almost certainly probing a column of much cooler dust along the line of sight and it may indeed reach dust at the sublimation temperature. A large range of spatial scales and temperatures are being merged into one aperture because of projection effects. It is thus difficult to draw any strong conclusions about the temperature in this feature without the $LM$-bands which should be more sensitive to hot dust. The $L$- and $M$-band fluxes measured using VLT/ISAAC by \citet{isbell2021} represent the AGN flux within 630 mas, and are certainly an upper limit on the $LM$ flux within the central aperture. However, if we perform a two-blackbody fit to the ISAAC $LM$ measurements in addition to our central aperture fluxes, we see that a very compact and extincted $1500$ K blackbody in addition to a larger $310$ K blackbody fit the data very well. So, it is possible that the central aperture contains dust at the sublimation temperature, but we cannot draw any strong conclusions without fully analyzing the $LM$ MATISSE data and spatially resolving the flux inside this central $\sim 10$ mas region. This will be done in a subsequent work.

\begin{figure}
    \centering
    \includegraphics[width=0.5\textwidth]{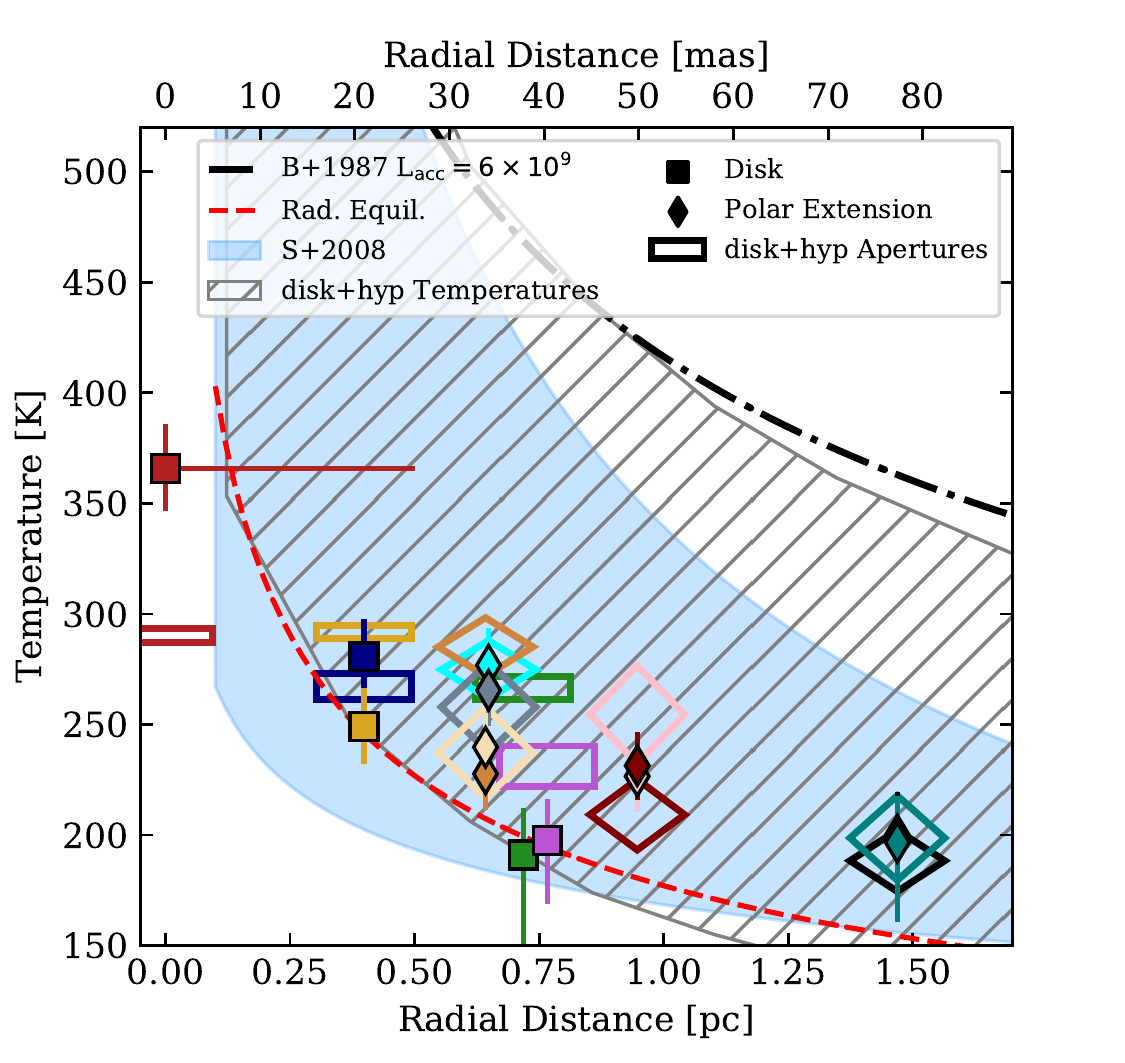}
    \caption{Radial temperature profile of the ``torus'' in Circinus based on 13 extracted apertures. Measurements along the disk-like structure in the center of image are marked with squares; measurements in the polar extensions are marked with diamonds. Colors correspond to the apertures given in Fig. \ref{fig:sed}. \textit{B+1987} is the radial profile from \citet{barvainis1987} and is given as the solid black line. The radial profile arising from simple radiation equilibrium is given as the dashed red line. In both analytic profiles the luminosity of the Circinus accretion disk is assumed to be $L_{\rm acc} = 6 \times 10^9 { L}_{\odot}$. 
    The shaded blue region labeled \textit{S+2008} represents the range of temperatures of the dust clumps at each radius in the standard clumpy torus model shown in \citet{schartmann2008}, Fig. 3. 
    The hashed region representes the range of temperatures of dust cells in the disk+hyp model.
    The boxes/diamonds represent fitted blackbody temperatures in our 13 apertures applied to disk+hyp models based on \citet{stalevski2019}; each box (for disk apertures) or diamond (for polar apertures) center is the median temperature $T$ and the height of each represents the range of temperatures fitted to the disk+hyp models. }
    \label{fig:profiles}
\end{figure}

\subsection{Central disk}
We find a thin disk-like structure along $\approx 30^{\circ}$. It is present in all wavelength channels, but most prominent at longer wavelengths.  In the continuum image, this disk is almost 1.9 pc in diameter and is unresolved in width. 
The extent of the disk is set by the $5\sigma$ contours at 12 and 12.7 \micron. 
Considering the dirty-beam (Appendix \ref{sec:appdb}), we expect artifacts in the form of secondary lobes at $\sim 100$ mas from the center along the disk PA, which indeed manifest themselves as low surface-brightness features near the edges of the images.  Nonetheless, the central part of the disk in our images has a high flux density and is robustly detected at S/N $\geq 5$. 

Evidence that the dust in this disk is relatively dense comes from the blackbody fits performed on the ``disk apertures.'' Here we see that the temperature falls quickly as one moves farther from the photo center; indeed, the lowest fitted temperatures in the image occur in the disk at a projected distance of only $0.7$ pc from the center. Taken together, the disk apertures (D0, D$_S$40, D$_N$40, D$_S$77, D$_N$72) exhibit a much steeper radial temperature gradient than apertures in the polar direction. 
The temperature profile of the disk is shown in both Fig. \ref{fig:profiles} and in the images themselves, as the disk becomes much less prominent at short wavelengths, indicating that the dust is relatively cool and the emission drops off significantly below $9$ \micron. The steep temperature gradient is possibly indicative of a dense environment wherein only the innermost dust has a direct view toward the accretion disk, and the outer clouds are heated only through re-radiation and photon scattering \citep[e.g.,][]{krolik2007}. 

The disk component places constraints on the inclination of the system. Assuming that the disk is both thin and axisymmetric with diameter 1.9 pc, the fact that we do not resolve the width of the disk ($\leq$ 9.5 mas $= 0.18$ pc at 12 \micron) indicates an inclination $i \gtrsim 83^{\circ}$. This is in agreement with the best disk+hyp model with $i\sim85^{\circ}$ matched to the closure phases and squared visibilities. This estimate can be considered a lower limit, as a more-realistic ``puffed-up'' disk would be thicker. $\gtrsim 83^{\circ}$ is closer to edge-on than on the galactic scale  \citep[$\sim 65^{\circ}$][]{freeman1977, elmouttie1998}, the ALMA CO(3-2) tilted ring estimation \citep[$i \gtrsim 70^{\circ}$][]{izumi2018}, and the T14 estimate for the MIR disk ($i > 75^{\circ}$). 
However, \citet{izumi2018} note that from 10 pc inward the inclination seems to increase, eventually reaching $i \sim 90^{\circ}$ for the warped H$_2$O maser disk \citep{greenhill2003}. The relatively dense dust of the MIR disk is consistent with the above and may lie in the same plane as the maser emission. The above assumes that the disk-like emission is the edge of the disk, rather than from reflected light on the top of a disk-like structure \citep[see e.g., T14;][]{stalevski2019}. We make this assumption because we observe no absorbing band on either side of the disk-emission.

The disk is aligned very well with the inner position angle of the H$_2$O maser emission \citep[$29 \pm 3^{\circ}$;][]{greenhill2003} as well as with the compact nuclear disk (CND) at 10s of pc found in ALMA CO(3-2) and [C{\sc{i}}](1-0) \citep[$32 \pm 1.9^{\circ}$;][]{izumi2018}. The entirety of the warped maser disk, moreover, fits within our $\leq 9.5$ mas $= 0.18$ pc thick dust disk. It is for this reason that we place the maser emission in the center of our disk; we do not have absolute astrometry from MATISSE, and so we must base the correspondence on the coincidence of PA and scale. Through Gaussian modeling, T14 also found a thin disk oriented along $46 \pm 3^{\circ}$ and with a FWHM of $1.1 \pm 0.3$ pc. The size of the disk in the T14 modeling is similar to what we measure.
The T14 disk orientation differs slightly from that of our imaged disk, but they $a)$ used differential phases rather than closure phases in their modeling; and $b)$ used Gaussian modeling which simplifies the structure and may combine components.
In \S \ref{sec:uvTest} we found that with the T14 \uv-coverage, our image disk could be oriented along $\approx$ $40^{\circ}$.

This dense disk of dust may play the role of the classical ``torus'', obscuring a direct view toward the BLR. However, we find two competing phenomena. First, we see in the central aperture that hot dust is present, and depending on the exact distribution of the dust in the $LM$ bands, we may even have a direct line of sight to dust at/near the sublimation temperature. This is, however, somewhat at odds with the steep temperature gradient we see across the disk. The thin disk must somehow be dense enough to shield some or most of the dust from directly seeing the sublimation zone or the central engine, but clumpy or low-density enough that we can see evidence of hot dust at or near the sublimation temperature. Authors such as \citet{kishimoto2011} and \citet{honig2012} hypothesize that a ``puffed-up'' inner region (a few sublimation radii in size) may act as the classical obscuring torus.

\subsection{Polar extension}

We find a large-scale structure oriented in the same direction in all wavelength channels. This structure is referred to as a ``polar extension'' because its primary axis lies perpendicular to the AGN orientation and along the radio jet. 
The polar extension in our imaging is a large ($\sim4 \times$ 1.5 pc) structure made up of warm ($> 200$ K) dust with major axis along $\approx -60^{\circ}$. This larger envelope contains significant substructure: most prominently enhanced brightness directly E and W of the disk center. The polar emission exhibits ``patchiness'' at a significance $\geq 3\sigma$ on scales similar to the beam size, most prominently in the 12.0 and 12.7 \micron~images. Patchiness in the image could arise from clumpy dust emission, though it is unlikely that we resolve individual clumps at this scale (10 mas = 0.19 pc). Nonetheless, these images provide direct evidence that the polar emission is not a smooth, continuous structure.

We find that the substructures of the polar emission exhibit spatial variation in temperature. 
At a similar projected distance, apertures E65 and W65 are marginally hotter than W64 and E64 ($\sim 270$ K vs $\sim 230$ K). Additionally, the dust comprising the polar extended regions remains warm ($\sim 200$ K) out to a projected distance of $\sim 1.5$ pc from the center of the structure. 
This behavior is significantly different than the dust temperature gradients along the disk, indicating differences in environment and density. The dust in the polar direction is likely less dense and/or more clumpy, as high temperatures at large distances require a relatively unobscured line of sight to the accretion disk.  
As shown in Fig. \ref{fig:profiles}, the temperatures in the polar emission are entirely consistent with predictions from radiative transfer modeling of clumpy media \citep[e.g.,][]{schartmann2008, stalevski2019}, however only the latter reproduce the interferometric observables. At much lower resolution, the MIR SEDs of nearby AGN have  shown that clumpy formalism is necessary to reproduce the relatively ``blue'' spectra indicative of an abundance of warm dust \citep[e.g.,][]{nenkova2008a, stalevski2016,honig2017}. 
At the parsec scale, our results support the predictions of clumpy models.

\subsubsection{E-W flux enhancements}
The morphology we recover is in accordance with previous single-dish $N$-band estimates of the polar dust, and with the MIDI results of T14. 
The primary position angle of the polar extension was estimated from VLT/VISIR observations to be $-80 \pm 10^{\circ}$ \citep[][]{asmus2016}. Similarly, the modeling done by T14 resulted in a $93_{-12}^{+6}$ mas FWHM ($\approx 2$ pc) Gaussian component with $T=304_{-8}^{+62}$ K and with a major axis along $-73 \pm 8$ deg. 
Both the single-dish PA and that of the large Gaussian component in T14 are directed more closely to E-W orientation than our imaging suggests. This is likely explained by the lack of resolution and the simplicity of the Gaussian modeling; the large structure in our imaging shows significant nonuniformities. Namely, enhancements in flux directly to the E and to the W of the image photocenter. If one considers a flux-weighted mean of the polar emission in our imaging, it would certainly be more similar to the PA$= 75\pm8^{\circ}$ as seen in T14. 
Indeed, the analysis by \citet{stalevski2017,stalevski2019} claims that the T14 large component is a simplified representation of an edge-brightened outflow cone, and they use this hypothesis to explain the discrepancy between the orientation of their polar outflow and the true pole of Circinus. 

We present two possible explanations for the bright E-W substructure of the polar emission.
The first is that the accretion disk in Circinus is tilted with respect to the central dust structures. If one considers that the central maser emission traces the orientation of the accretion disk (supported by the agreement with the radio jet position angle, assuming the jets originate in the central region), then one can relatively simply explain the asymmetric illumination of the polar extension. We show in Fig. \ref{fig:continuum} a line tracing the radio jet orientation. This line touches both the E and W flux-enhanced regions of the image. Due to the anisotropic nature of accretion disk emission, any dust some angle $\theta$ away from the ``face'' of the accretion disk is illuminated by a factor $\propto \cos\theta (2\cos\theta +1)$ less than the dust which does see the ``face'' \citep{netzer1987}. The features we observe end more abruptly than this function suggests, but this could be due to patchiness or clumpiness of the dust. The idea of an accretion disk tilted with respect to the large-scale structures in Circinus is not new. \citet{greenhill2003} suggests that the orientation of the accretion disk should only be ``weakly coupled via gravity to the surrounding large-scale dynamical structures'' because the central engine has a sphere of influence with a radius of only a few pc \citep{curran1998}. Using VISIR images, MIDI observations, and the SED of Circinus, T14 as well as \citet{stalevski2017,stalevski2019} hypothesized that a warped or tilted accretion disk \citep[as described by e.g.,][]{petterson1977,nayakshin2005} was required to asymmetrically illuminate the polar dust in their modeling. Hydrodynamic modeling of the central structures by \citet{wada2012} predicts that symmetric radiation-driven outflow cones should form perpendicular to the accretion disk. So while our observations suggest that the illumination of the polar dust is asymmetric --possibly from a tilted accretion disk-- we cannot at this time explain why or how such a tilt occurred. 
The second possibility is that there is simply more material along the E-W direction; indeed \citet{greenhill2003} speculated that the warped accretion disk could channel material in the nuclear outflow. This hypothesis is in better agreement with the \citet{wada2012} modeling, as in this case the polar outflows would be symmetric w.r.t. the accretion disk. The higher temperatures of the E-W flux-enhancements with respect to the apertures at the same projected distance argue in favor of the direct-illumination hypothesis. An overdensity of material should exhibit cooler temperatures due to dust self-shielding (as seen in the disk). This is merely a qualitative agreement, and in order to distinguish between these two hypotheses, detailed modeling of the formation of the outflow cones in the presence of a warped accretion disk will be crucial.

\subsubsection{Connection to larger scales}
It is clear in the MATISSE imaging that the majority of thermal dust emission in the center of Circinus comes from the polar extension, but its full extent is poorly constrained.  
In our imaging, any structure larger than those probed by the shortest baselines is resolved out; this means for imaging using the MIDI AT baselines we are not sensitive to structures larger than 688 mas at 12 \micron. 
This is strictly an upper limit, however, and in the image reconstruction process we $a)$ limit the FOV to 600 mas, and $b)$ apply an object mask with a radius $160$ mas. The object masking heavily suppresses any structure which falls outside of the specified radius. 
We can, nonetheless, confidently state that there is $N$-band emission out to $\sim 1.5$ pc from the center to both the NW and SE, and that the emission shows a flux enhancement to the E and W of the image center. 

The VISIR-SAM data were fit in the image plane with a Gaussian having FWHM $3.3 \times 2.2$ pc and major axis along $PA = 72^{\circ}$. This is larger than either the MATISSE images or the MIDI modeling (with FWHM = 2 pc), indicating that the MATISSE images do not capture the true extent of the structure. The position angle of the SAM data matches the T14 result, though it is likely also flux-biased toward the South due to the E-W flux enhancements. 
Continuing to lower resolution, the $N$-band VLT/VISIR images in \citet{asmus2014, asmus2016} show that in Circinus roughly 60\% of the flux is extended farther than 5.24 pc and at PA $= 100 \pm 10^{\circ}$. It is clear that the polar structures we see in our images extend continuously outward past 5 pc.

\subsection{Overall morphology}
We present the first model-independent image of the circumnuclear dust in Circinus. The recovered combination of a geometrically thin disk and large-scale polar emission supports previous MIR interferometric findings, but newly imaged substructures hint at complexity unmatched in existing modeling. In particular, we find that the disk is simultaneously dense and yet allows emission from hot dust to radiate through; we find that an unresolved component lies 10 mas NE of the photocenter along the disk; and we find significant flux enhancements in the polar emission E and W of the disk-center. 

The size of circumnuclear dust structures has been shown to vary with AGN luminosity \citep[e.g.,][]{kishimoto2011,burtscher2013}. The scales measured herein of the circumnuclear structures in Circinus --namely a thin disk with diameter 1.9 pc and $\gtrsim 4$ pc polar emission-- with $L_{\rm AGN} = 6\times 10^9 - 7\times 10^{10}~{L}_{\odot}$ \citep{arevalo2014,ricci2015,tristram2014,oliva1999} place a constraint on the luminosity-dependent scaling of the dust structures in AGN. \citet{leftley2019} showed that the ratio of extended flux to unresolved flux increased with Eddington ratio ($\epsilon_{\rm Edd}$), claiming that this implied the presence of more dust in a radiation-driven wind for a higher $\epsilon_{\rm Edd}$. 
Circinus, with $\epsilon_{\rm Edd} \sim 0.2$ \citep{greenhill2003}, is dominated by polar dust emission. 
We measure the flux of the unresolved component to be $F_{{\rm pt,} 12\mu{\rm m}} = 0.77\pm 0.04$ Jy, which is $6.3 \pm 0.3$\% of the total flux at 12 \micron. At 8.9 \micron~we measure $F_{{\rm pt,} 8.9\mu{\rm m}} = 0.39\pm0.01$ Jy, which is $9.0\pm0.2$\% of the total flux. 
The fraction at 12 \micron~is significantly smaller than previously reported \citep[20\% and 10\% at 12 \micron~in][respectively]{leftley2019,lopez-gonzaga2016}, but they relied on simple two-Gaussian modeling of MIDI data.

Disk+wind radiative-transfer models \citep{honig2017, stalevski2019} have recently been invoked to explain the polar emission found in a number of nearby AGN \citep[e.g.,][]{tristram2007, burtscher2013,tristram2014,lopez-gonzaga2014,lopez-gonzaga2016,leftley2018}. Fits to the NIR and MIR SEDs of nearby AGN have shown that disk+wind models provide the best match to the overall SED, reproducing the MIR flux through large-scale emission and NIR flux via reflected light from the accretion disk in the windy outflow \citep[e.g.,][]{martinez-paredes2020, isbell2021}. The disk+wind morphology in the radiative-transfer models is supported by hydrodynamical and radiation-hydro modeling \citep{wada2012,wada2016,williamson2020,venanzi2020}, but has had few direct observational constraints. The images presented in this work, with a thin disk (1.9 pc $\times \leq 0.18$ pc) and polar emission ($\sim 4 \times 1.5$ pc) perpendicular to it, resemble the disk+wind models only in broad strokes. Modifications to the disk+wind model in \citet{stalevski2019} explain the E-W flux enhancements in the polar emission via a tilted accretion disk, but the dynamical stability of such a shift in radiation pressure remains untested. Hydrodynamical models produce structures symmetric about the accretion disk \citep[][]{wada2012, venanzi2020}, so tilting with respect to the dusty structures may play a larger role. Whether this is specific to Circinus or a more general feature remains to be explored.

Only one other AGN has been imaged with MATISSE so far: NGC\,1068. Imaging work by \citet{gamezrosas2022} has revealed a quite different circumnuclear dust morphology than we recover. In NGC\,1068 at 12 \micron, they find a disk-like structure $\sim 2$ pc in diameter with emission extending nearly perpendicular to it, similar to what we see with the disk and E-W polar flux enhancements. 
However, the NGC\,1068 and Circinus morphologies differ significantly at other wavelengths. At 8.5 \micron~and in the $LM$-bands, the NGC\,1068 emission is resolved into a ring-like structure with $720$ K dust embedded within. We have shown that hot dust can make up the unresolved flux in Circinus, and
the $LM$ data can help clarify the situation, as they probe the $\gtrsim 500$ K dust morphology. Finally, \citet{jaffe2004,lopez-gonzaga2014}, and \citet{gamezrosas2022} showed that in NGC\,1068, the standard ISM dust we use does not reproduce the observed SEDs. The effects of varying dust composition will be explored in future work.

Future $N$-band observations with the MATISSE ATs will yield the first closure phase measurements of the $\gtrsim 1$ pc dust, further improving our imaging capabilites beyond the MIDI data. In a subsequent paper, we will utilize the $LM$-band MATISSE data, with $\sim 3$ mas resolution, to probe the hotter dust at small scales both within the disk and perhaps at the origins of the polar extension. We show that the central aperture can contain dust near the sublimation temperature, and a detailed study of the $LM$-data can give insights into the density, thickness, and perhaps the clumpiness of the disk. Circinus and NGC\,1068 are laboratories in which to study the circumnuclear dust at extremely small physical scales, but the ongoing the MATISSE AGN Programme aims toward a statistical understanding of the central dust scaling and relation to the SMBH.

\section{Conclusions}
\label{sec:conc}

In this work we present the first images of the circumnuclear dust in the Seyfert 2 galaxy Circinus. These images were reconstructed with IRBis using 150 correlated fluxes and 100 closure phases in the $N$-band from VLTI/MATISSE. Closure phase measurements of Circinus are reported here, and their novel inclusion in MATISSE observations makes imaging possible for the first time. The above results are largely in agreement with previous observations from MIDI \citep[][]{tristram2007,tristram2014} and VISIR \citep{asmus2014}. But our images, moreover, are model-independent and show new substructure which can be used to further constrain physical modeling of circumnuclear dust in AGN. Through analysis of the interferometric observables and the images reconstructed in seven independent wavelength channels we

\begin{enumerate}
    \item Show that correlated flux measurements on individual baselines have not changed over the last 17 years, implying that the underlying structures remain unchanged from the MIDI observations obtained between 2004 and 2011.
    
    \item Find significant substructure in the circumnuclear dust. The circumnuclear dust can be separated into several components: central, unresolved flux; a thin disk 1.9 pc in diameter; polar emission ($\sim 4\times 1.5$ pc) extending orthogonal to the disk and exhibiting patchiness; and flux enhancements E and W of the disk embedded within the polar dust.

    \item Report that the polar dust makes up $\sim 60\%$ of the total flux, increasing toward longer wavelengths. The unresolved flux makes up $\lesssim 10\%$, increasing toward shorter wavelengths and further hinting at the presence of hot dust.
    
    \item Measure SEDs in 13 apertures across the structures and fit temperature and extinction values to blackbodies in those apertures. We fit hotter dust temperatures ($T=367_{-26}^{+30}$ K) in the central aperture along with warm dust ($T\gtrsim 200$ K) $1.5$ pc from the center, indicating a clumpy circumnuclear medium. 
    We clearly distinguish the radial temperature profiles of the disk and the polar extension: the disk shows a steeper temperature gradient indicating dense material; the polar emission shows a much flatter temperature profile with warm temperatures out to $2$ pc from the center.

    \item Recover a remarkably symmetric object, in terms of both flux and temperature distributions. We fit A$_{\rm V} = 28.5_{-7.7}^{+8.5}$ mag, consistent with the galactic-scale value \citep[A$_{\rm V} = 28\pm7$][]{wilson2000}. 
    We find no evidence of an absorption gradient across the field, in contrast to previous results \citep[i.e.,][]{tristram2014}. Our new results indicate the presence of a foreground dust screen with very little local variation.
    
    \item See that on large scales, the recovered morphology of the $N$-band dust in Circinus resembles the results of disk+wind modeling \citep[e.g.,][]{wada2016,stalevski2019}, but new questions are raised because the subparsec dust is imaged here for the first time. 
    We find that the temperature distribution is well-reproduced by the clumpy torus models of \citet{schartmann2008} and \citet{ stalevski2019}. The \citet{schartmann2008} models do not, however, match the imaged morphology. The disk+hyp models better match the structure, but discrepancies are found in the central and disk apertures, indicating modifications to the disk component are necessary in the models. 
    Using a suite of disk+hyp models based on \citet{stalevski2019}, we find that a large range of clump densities and disk filling-factors can match the data within the uncertainties of the images and interferometric observables.
    
    \item Discover inhomogeneities in the polar dust emission: namely significant patchiness on scales of the resolution element; and flux enhancements directly to the E and W of the disk. The here-discovered patchiness is the first direct evidence that the polar dust is not a smooth, continuous structure but is rather clumpy. The E-W flux enhancement raises questions about the relation of the accretion disk to the larger dust structures.

\end{enumerate}

The imaged substructures  and temperature distributions presented herein serve as a direct constraint on future physical modeling of the circumnuclear dust. In the disk+wind model, the thin disk we image is related to inflowing material, while the polar emission represents a radiation-driven outflow. How these components relate to large-scale ($\gtrsim 10$ pc) structures and furthermore to the host galaxy can be tested in both hydrodynamical modeling and future observations, specifically with the MATISSE ATs. It is clear that the classic geometrically-thick torus is not present in our imaging, but the (nearly) rotationally symmetric structures we recover can play much the same role; yet the detailed implications for AGN Unification remain to be explored through modeling and the MATISSE AGN Programme.

\begin{acknowledgements}
We thank the VLTI Paranal staff for all their help and good company during observations, the entire MATISSE Consortium for their support, and the anonymous referee for their helpful comments which have improved the manuscript. 

J.W.I. thanks Felix Bosco and Thomas Jackson for insightful discussions.

J.S.B. acknowledges the full support from the UNAM PAPIIT project IA 101220 and from the CONACyT “Ciencia de Frontera” project 263975.

M.S. acknowledges support by the Ministry of Education, Science and Technological Development of the Republic of Serbia through the contract no. 451-03-9/2022-14/200002 and by the Science Fund of the Republic of Serbia, PROMIS 6060916, BOWIE.

This research has made extensive use of NASA’s Astrophysics Data System; the SIMBAD database and VizieR catalogue access tool, operated at CDS, Strasbourg, France; and the python packages \pack{astropy}, \pack{emcee}, \pack{scipy}, and \pack{matplotlib}.
\end{acknowledgements}

\bibliographystyle{aa}
\bibliography{mybib}

\begin{appendix}
\section{MATISSE correlated fluxes}
In Figs. \ref{fig:cflux0}-\ref{fig:cflux3} we present the correlated flux for each $\uv$-point, reduced and calibrated as described in \S\ref{sec:obs}. The total photometric flux (``the zero-baseline flux'') is included in the first panel of Fig. \ref{fig:cflux0}.

\begin{figure*}

    \includegraphics[width=.98\textwidth]{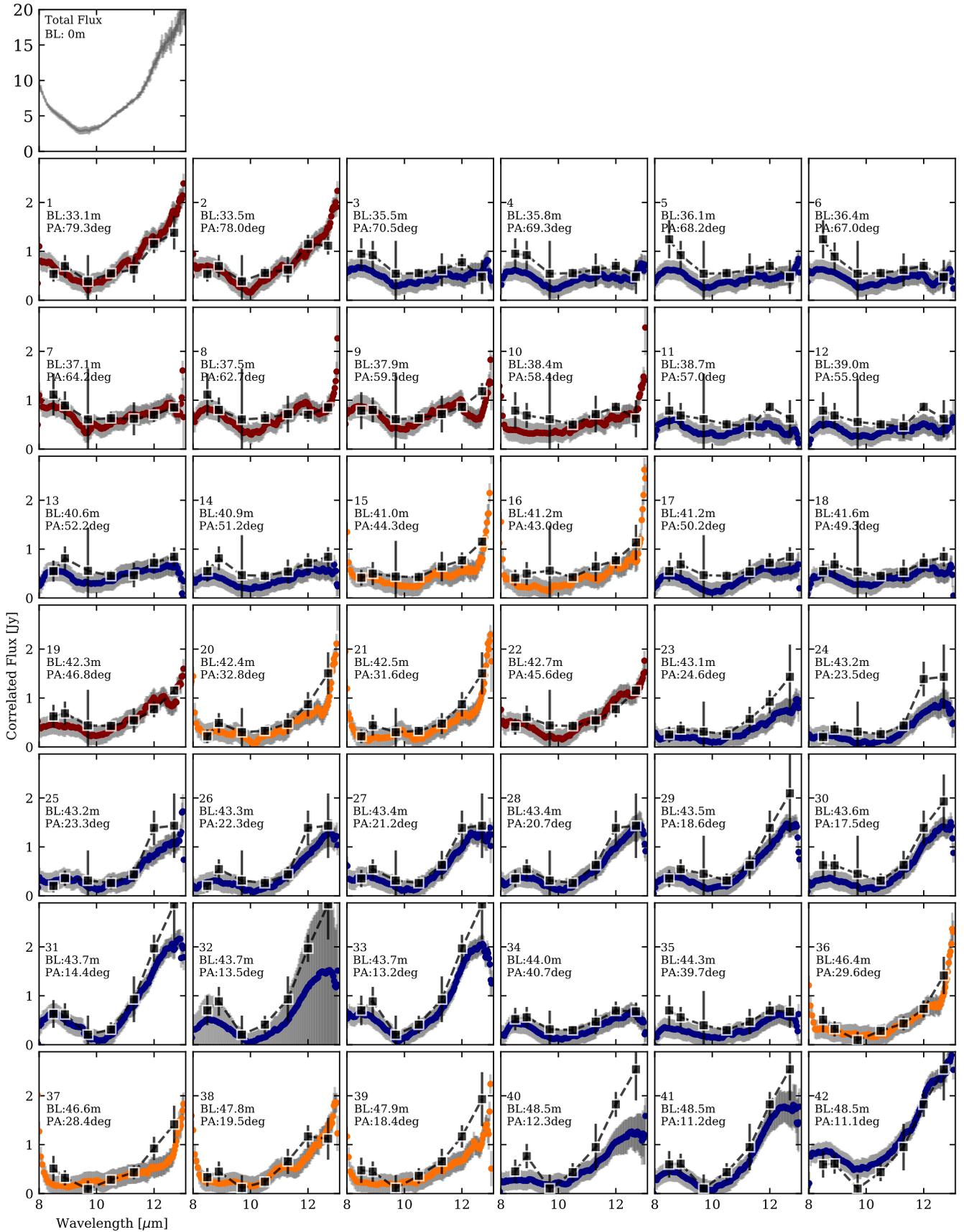}
    \caption{\small{Correlated flux data for Circinus from March 2020 (blue), February 2021 (yellow), and May 2021 (red). The black points are simulated values extracted from the final images, with errors estimated using the $1\sigma$ error maps (described in \S \ref{sec:imerrors}). The total photometric flux is included in the first panel. Presented errors come from both the calibrator flux uncertainty and the statistical variation of the observables within a set of observing cycles. Near 8 and 13 \micron~one can see flux variations due to the edges of the atmospheric window.} }
    \label{fig:cflux0}
\end{figure*}
\begin{figure*}
    \centering
    \ContinuedFloat
    \includegraphics[width=1.0\textwidth]{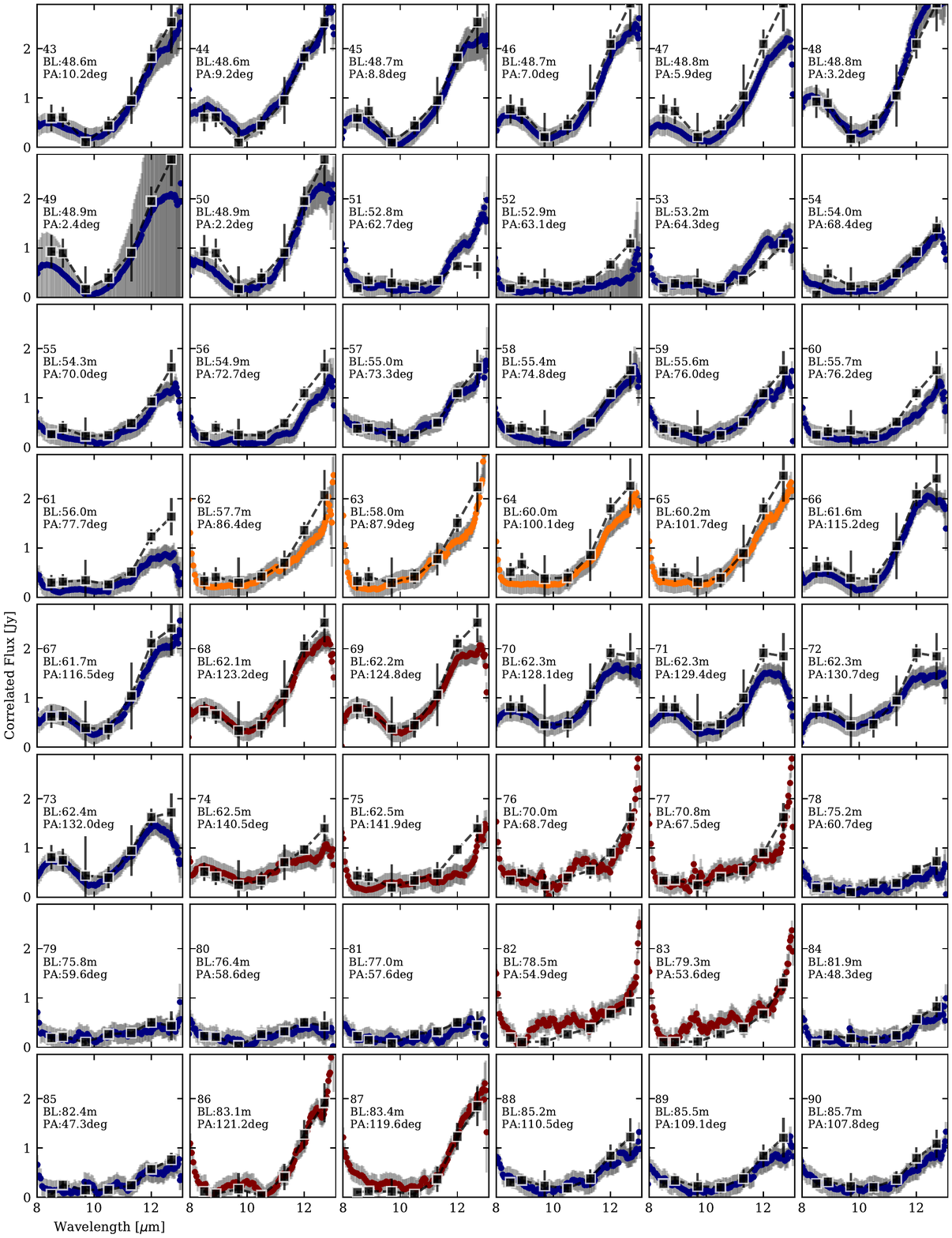}
    \caption{continued.}
    \label{fig:cflux1}
\end{figure*}
\begin{figure*}
    \centering
    \ContinuedFloat
    \includegraphics[width=1.0\textwidth]{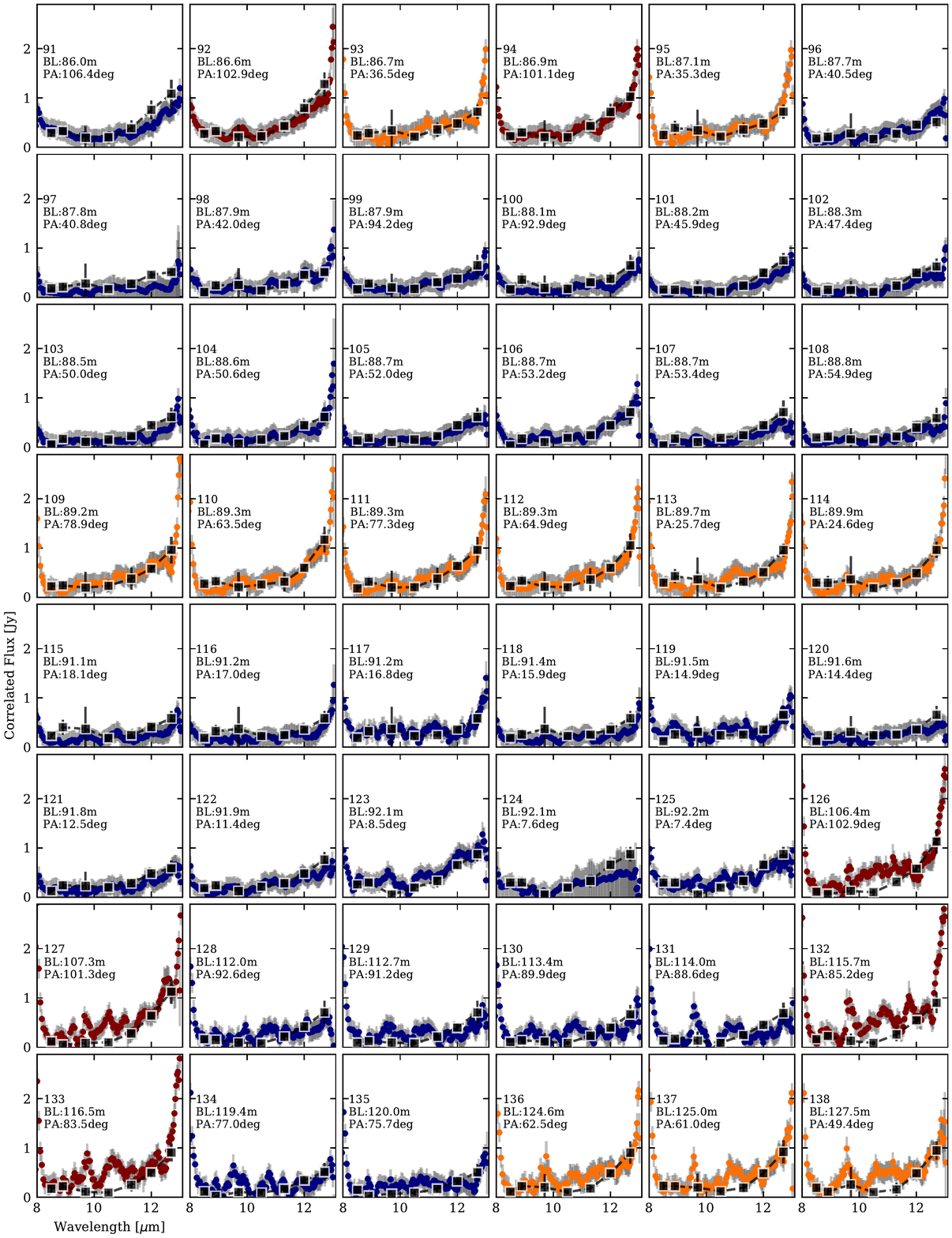}
    \caption{continued.}
    \label{fig:cflux2}
\end{figure*}
\begin{figure*}
    \centering
    \ContinuedFloat
    \includegraphics[width=1.0\textwidth]{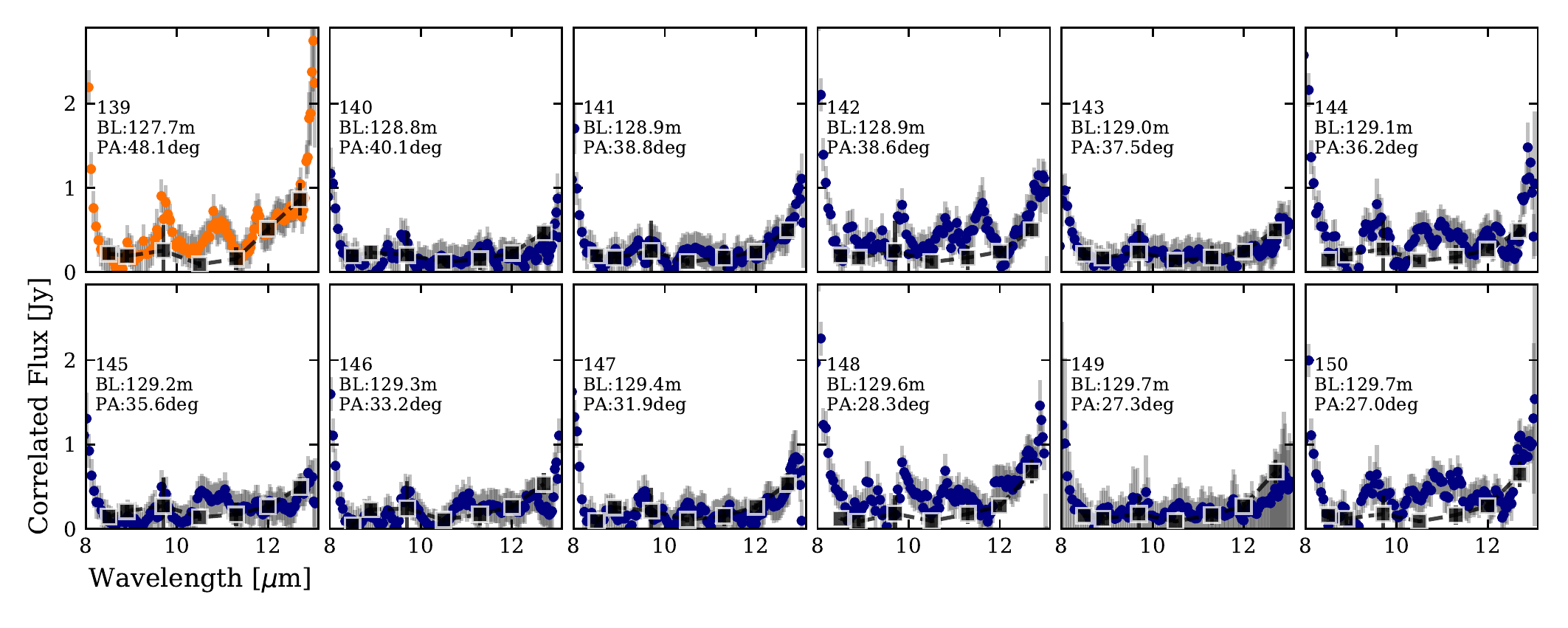}
    \caption{continued.}
    \label{fig:cflux3}
\end{figure*}

\section{MATISSE closure phases}
In Figs. \ref{fig:t3phi0}-\ref{fig:t3phi3} we present the closure phase for each closure triangle, reduced and calibrated as described in \S\ref{sec:obs}.

\begin{figure*}
    \centering
    \includegraphics[width=0.98\textwidth]{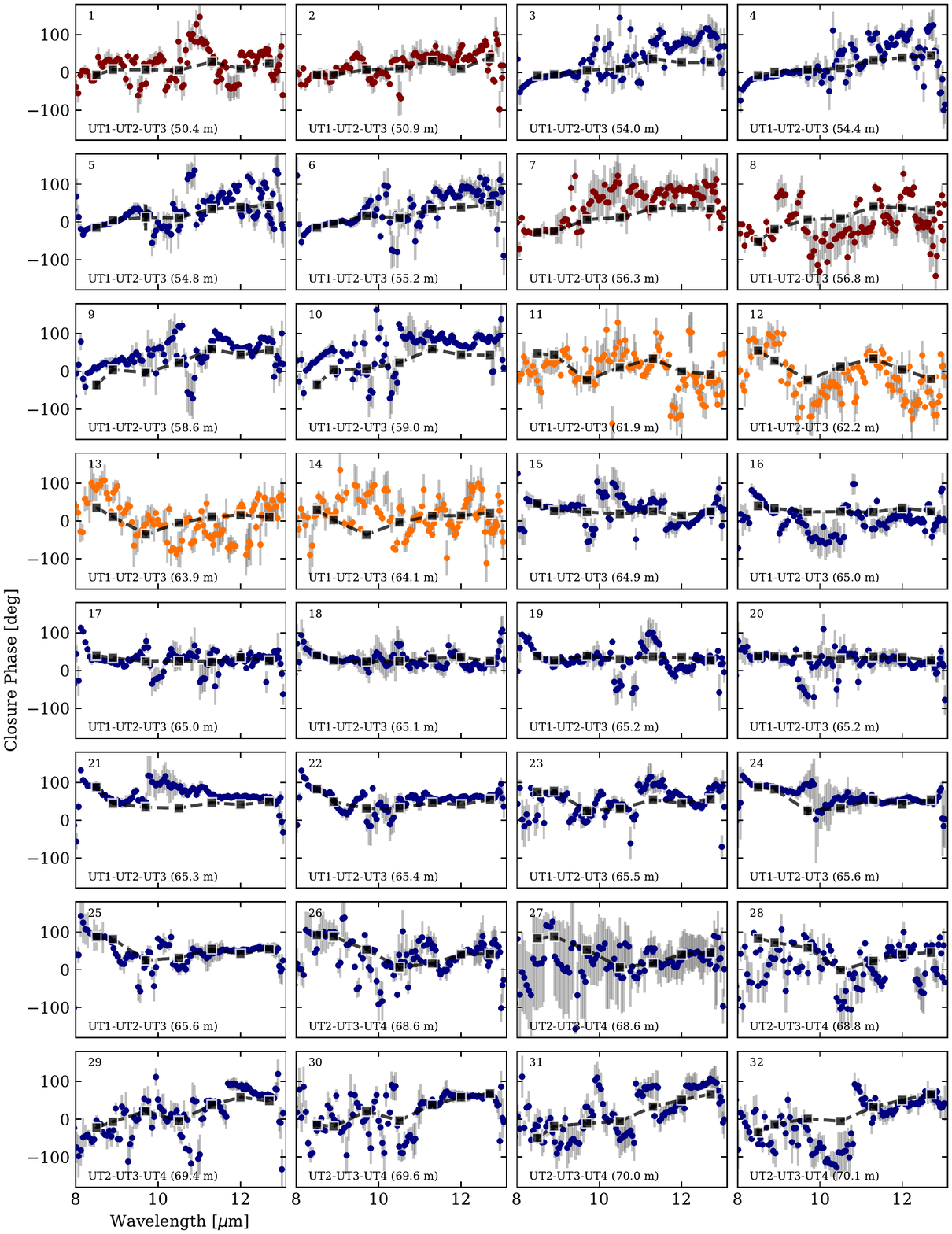}
    
    \caption{\small{Closure phase data for Circinus from March 2020 (blue), February 2021 (yellow), and May 2021 (red). Presented errors come from both the calibrator phase uncertainty and the statistical variation of the observables within a set of observing cycles. The black points are simulated values extracted from the final images, with errors estimated using the $1\sigma$ error maps (described in \S \ref{sec:imerrors}). The panels are sorted by length of the longest projected baseline in the closure triangle. 
    }
    }
    \label{fig:t3phi0}
\end{figure*}
\begin{figure*}
    \ContinuedFloat
    \centering
    \includegraphics[width=1.0\textwidth]{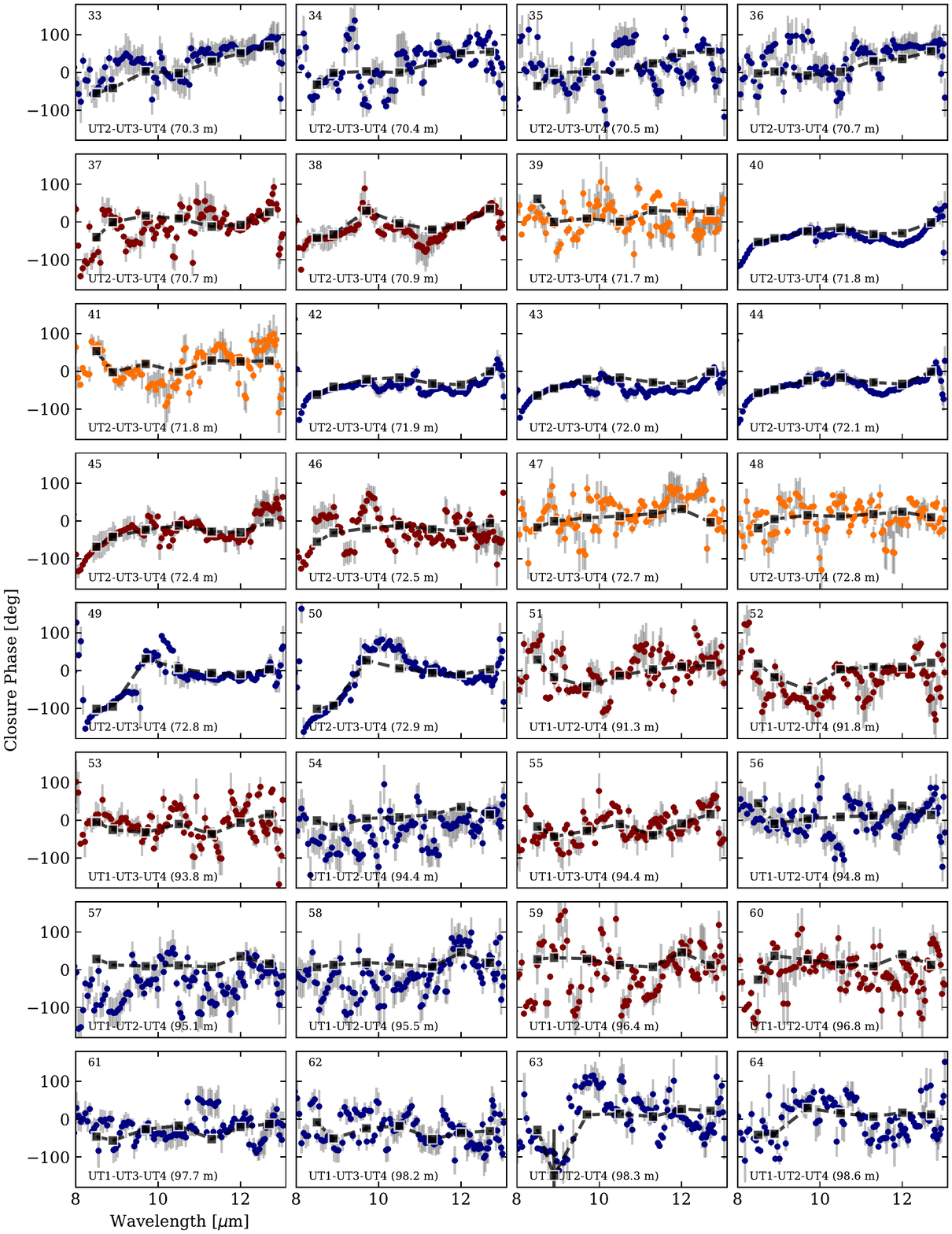}
    \caption{continued.}
    \label{fig:t3phi1}
\end{figure*}
\begin{figure*}
\ContinuedFloat
    \centering
    \includegraphics[width=1.0\textwidth]{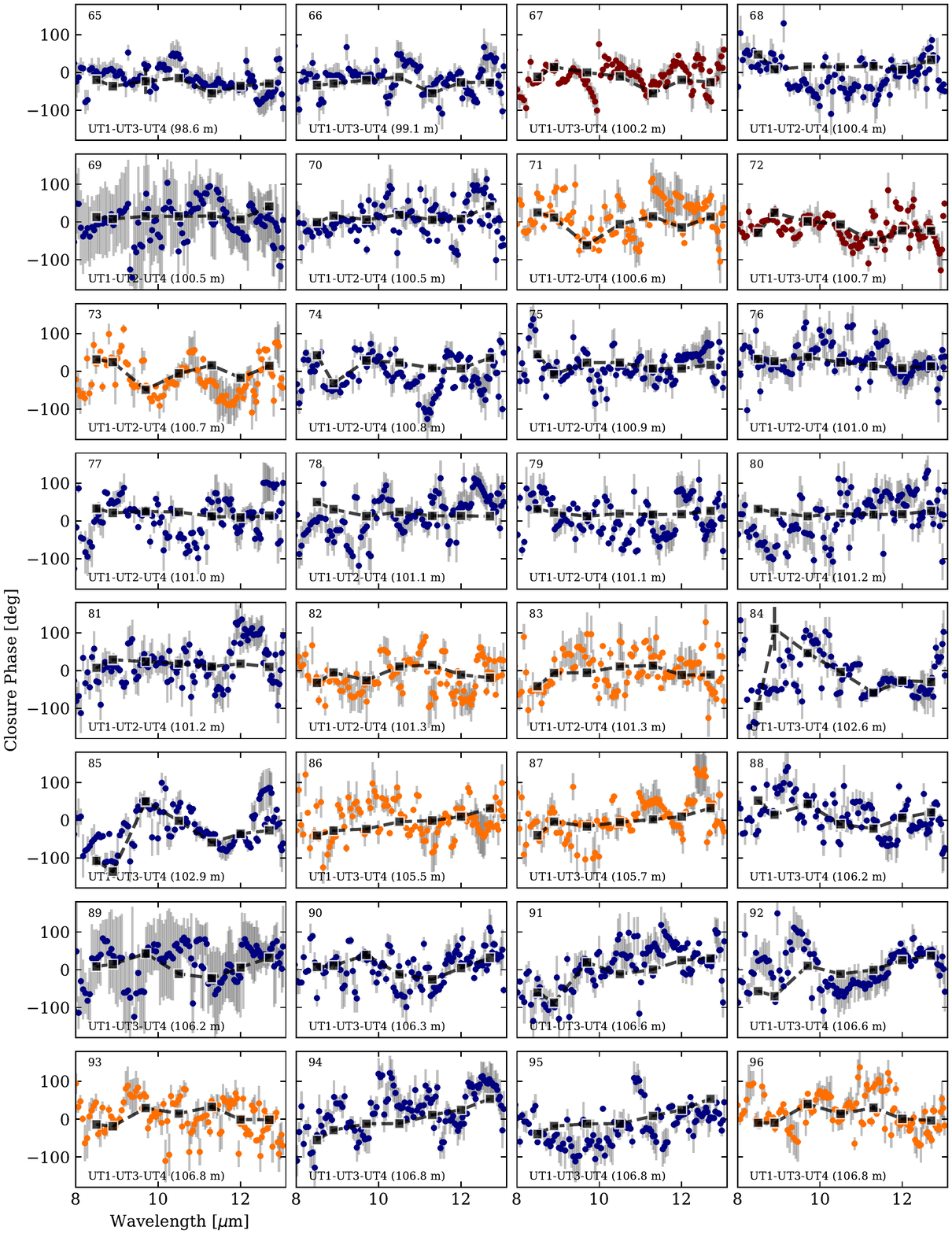}
    \caption{continued.}
    \label{fig:t3phi2}
\end{figure*}
\begin{figure*}
\ContinuedFloat
    \centering
    \includegraphics[width=1.0\textwidth]{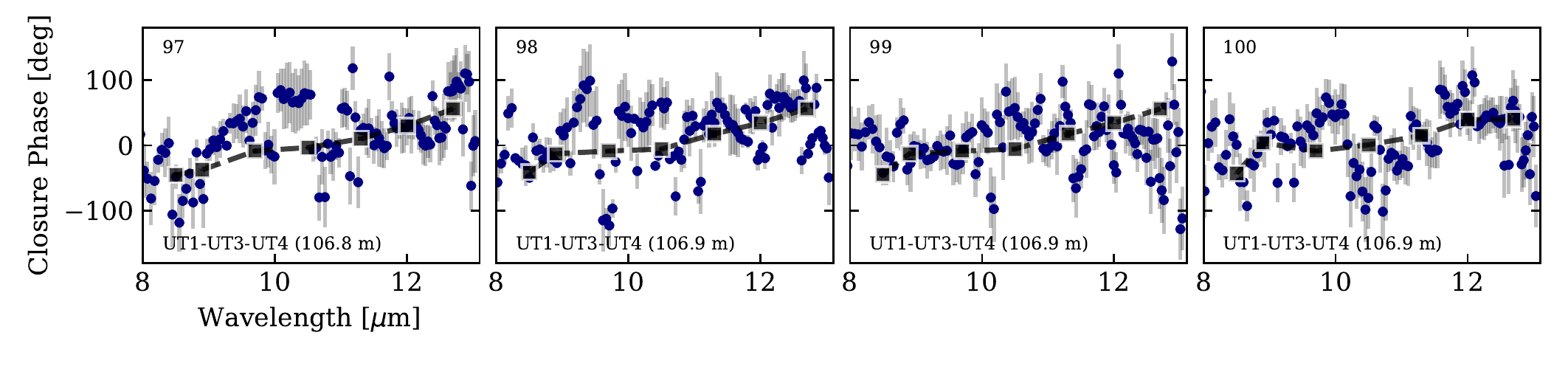}
    \caption{continued.}
    \label{fig:t3phi3}
\end{figure*}

\section{VISIR-SAM data}
\begin{figure*}
    \centering
    \includegraphics[width=0.85\textwidth]{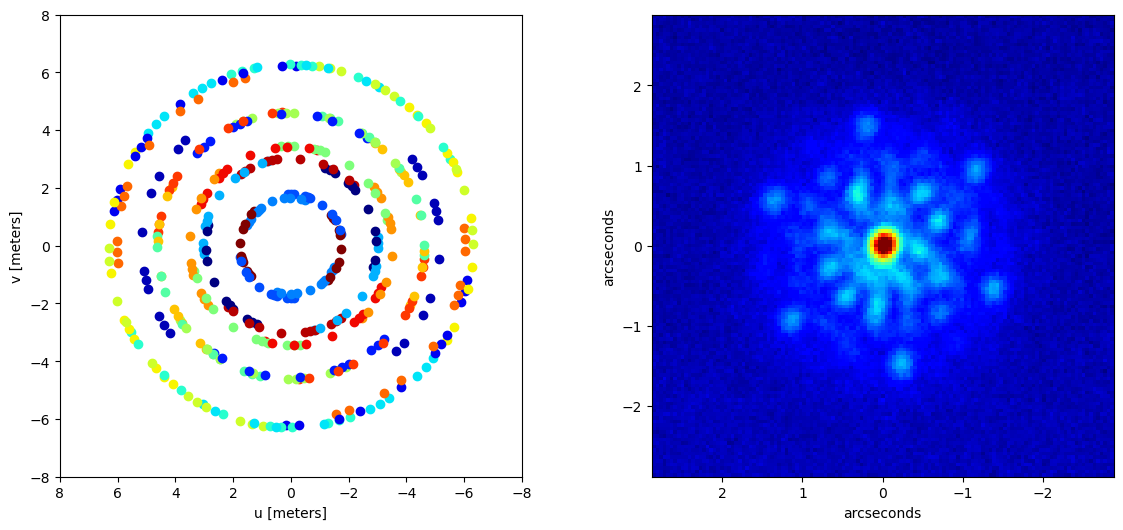}
    \caption{u-v coverage obtained with the VISIR-SAM data (\textit{left}), the different colors indicate the 21 different baselines in the data. Snapshot of the Circinus interferogram obtained with the VISIR-SAM data (\textit{right}).}
    \label{fig:visir_uv}
\end{figure*}

\begin{figure*}
    \centering
    \includegraphics[width=0.85\textwidth]{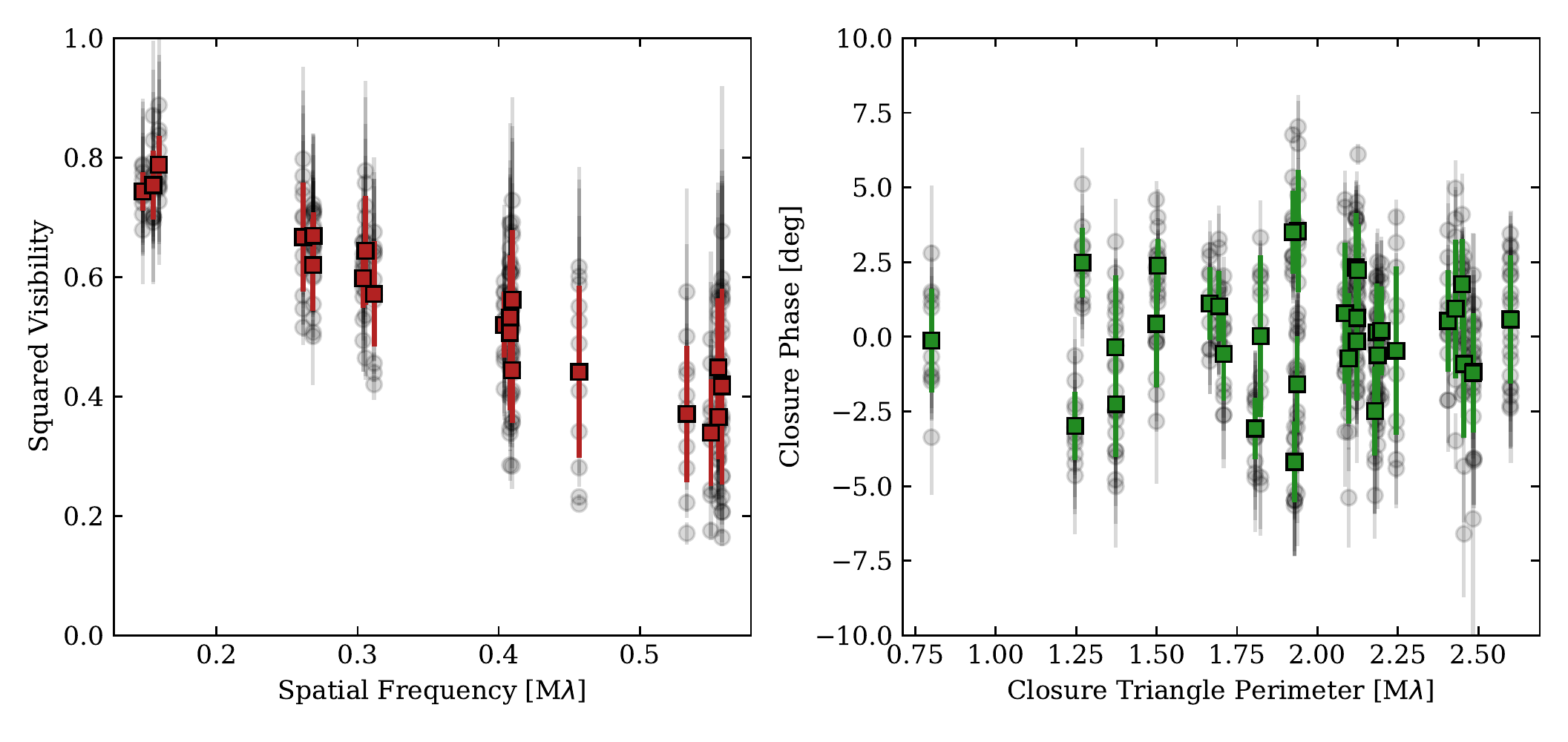}
    \caption{VISIR-SAM data used in the diagnosis of large-scale closure phases. In the (\textit{left}) panel we show squared visibilities. The individual observations are in gray, and the mean value over the cycles at a given baseline is in red with $1\sigma$ error bars coming from the standard deviation of the cycles. In the (\textit{right}) panel we show the same for closure phase, with mean values in green. }
    \label{fig:visirsam}
\end{figure*}

In Figs \ref{fig:visir_uv} and \ref{fig:visirsam} we show the measured VISIR-SAM data from the reduction as described in \S\ref{sec:visirsam}.

\section{Dirty beam}
\label{sec:appdb}
\begin{figure*}
    \centering
    \includegraphics[width=0.75\textwidth]{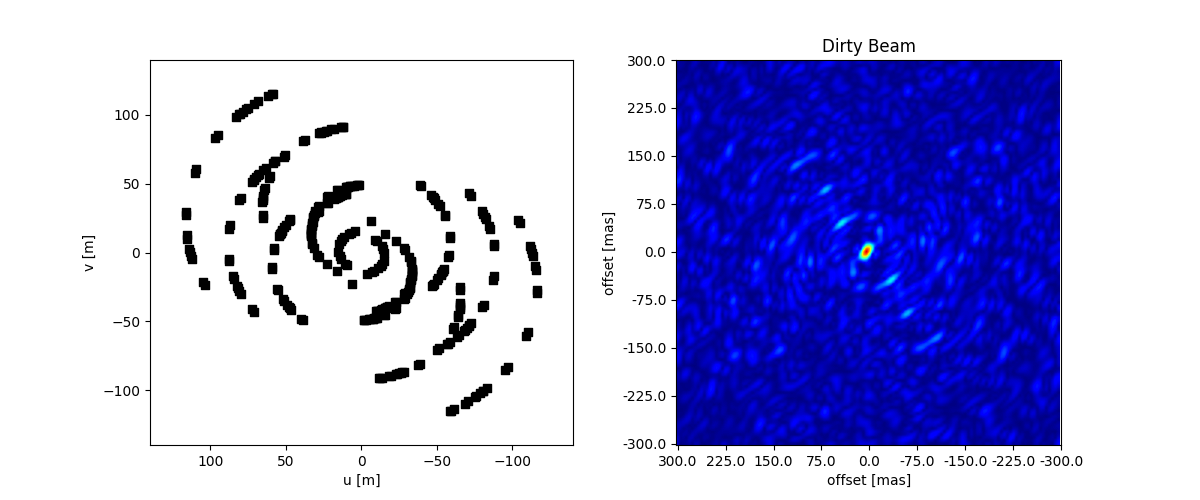}
    \caption{Dirty beam estimated for the combined MIDI AT and MATISSE UT \uv-coverage. On the (\textit{left}) we show the final \uv-coverage, and on the (\textit{right}) we show the resulting dirty beam with square-root scaling inside a 600 mas window.  }
    \label{fig:dirtyBeam}
\end{figure*}

We estimate the dirty beam in the typical way in order to identify image artifacts. In the \uv-plane, we set the squared visibility at each \uv-point we observed ($\pm 4.05$ m, the UT radius) to 1 and the surrounding points to 0. 
We set the phase to 0 deg across the \uv-plane. We finally take the inverse Fourier transform of this complex array to obtain an estimate of the dirty beam (shown in Fig. \ref{fig:dirtyBeam}).

\section{Imaging with and without ATs}
\label{sec:appImage}
\begin{figure*}
    \centering
    \includegraphics[width=0.9\textwidth]{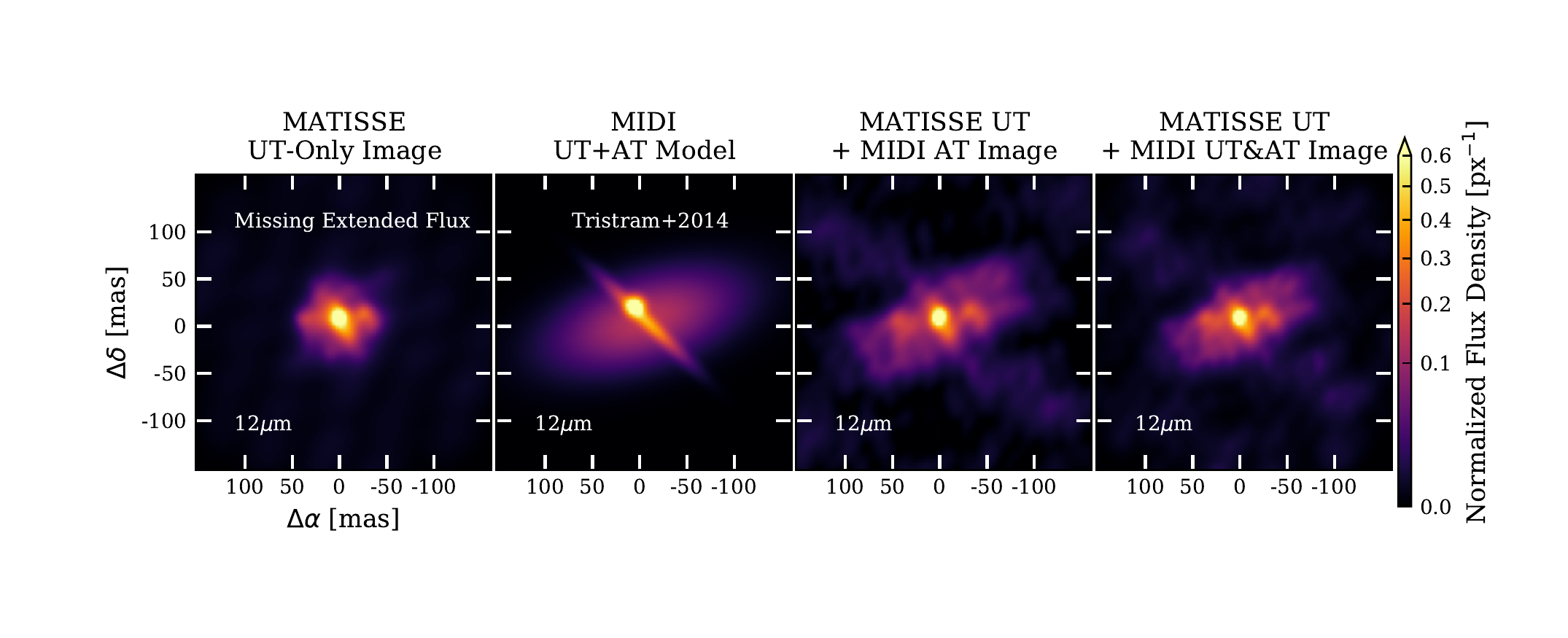}
    \caption{Comparison of image reconstruction using different \uv-samplings. In the (\textit{leftmost panel}) we show the image resulting from MATISSE UT \uv-coverage alone. In the (\textit{second panel}) we show the MIDI UT+AT Gaussian model from \citet{tristram2014}. In the (\textit{third panel})  we show the image reconstruction resulting from the combination of MATISSE UT and MIDI AT \uv-coverage. In the (\textit{rightmost panel}) we show the results of imaging using the MATISSE UT, the MIDI AT, and the MIDI UT data, with closure phases in the MIDI data set by the T14 Gaussian model. The interior structures (a disk, an unresolved source, and bright E-W flux enhancements) are present in all image reconstructions, implying their fidelity. }
    \label{fig:app_utonly}
\end{figure*}

In Fig. \ref{fig:app_utonly} we show the effects of imaging with and without the MIDI AT baselines. We stress that the MIDI AT baseline inclusion is necessary due to the resolved nature of this AGN, as shown in both MIR interferometric and single-dish observations. The MIDI AT baselines require the synthesis of closure phase triangles in order to match the IRBis formatting. We set the closure phases involving these baselines to $0\pm 180^{\circ}$, such that they do not bias the imaging.  We justify the inclusion of these baselines through the following arguments:
    First, the correlated flux values for all 30 MIDI \uv-points within 4m of a MATISSE point show $<2\sigma$ variation over 10 years (\S \ref{sec:fluxvar}).
    Second, the AT baselines from MIDI transition continuously to the MATISSE UT baselines around 30m (i.e., variations within the 0.2 Jy correlated flux uncertainties).
    Finally,  VISIR-SAM imaging of Circinus shows $0.1 \pm 2.5^{\circ}$ closure phases on $\leq 6.3$ m baselines (\S \ref{sec:visirsam}). This agrees with T14's Gaussian modeling of the MIDI data which gives $\approx 0^{\circ}$ closure phases for baselines $\leq 30$ m.

Nonetheless, it is instructive to see which structures arise as a result of the MATISSE-only imaging. We show the 12 \micron~UT-only reconstruction in Fig. \ref{fig:app_utonly} alongside the Gaussian model of T14 which used both UTs and ATs from MIDI (without closure phases) and the 12 \micron~image reconstruction as detailed in \S \ref{sec:imarec}. We see that the central $\approx 1$ pc is nearly identical in the two images, and notably the bright features E-W of the center remain prominent in both setups. The disk-like component is perhaps even more obvious in the UT-only image, given the same color scaling. The largest difference between the images is the lack of large-scale extended flux, but this is expected as the UTs shortest baseline corresponds to $\approx 40$ mas at 12 \micron, and structures larger than this are suppressed. 

\section{Image error estimates}
\label{sec:appErrors}
\begin{figure*}
    \centering
    \includegraphics[width=0.9\textwidth]{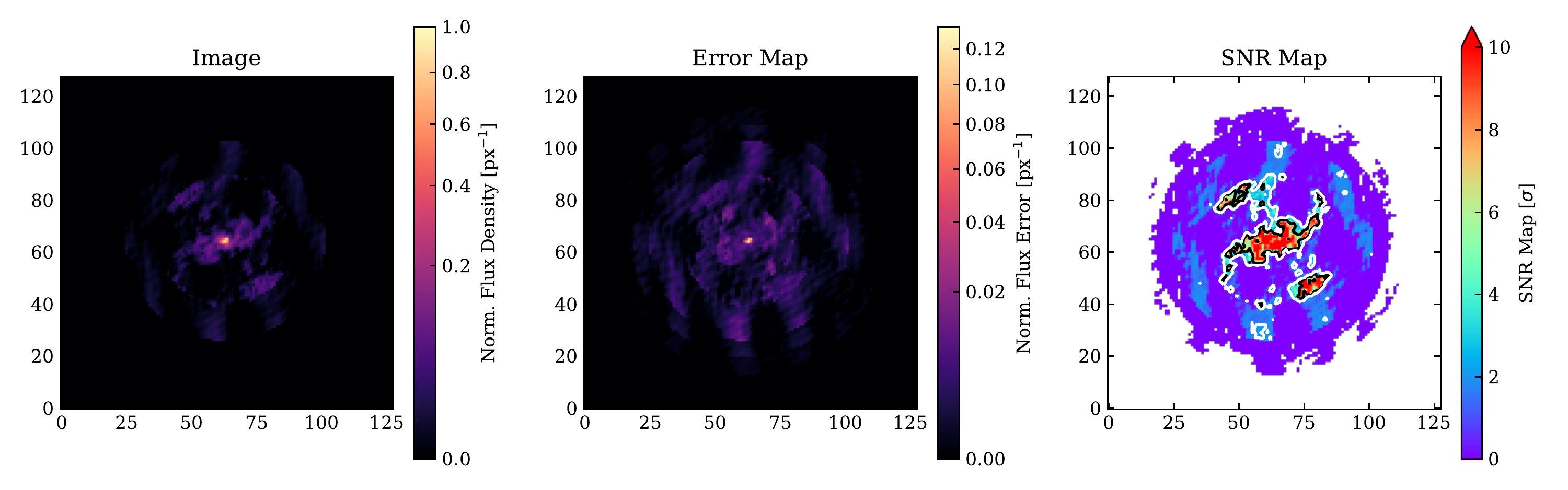}
    \caption{Final image reconstructions and error estimates. From left to right: the median image, error map, and S/N map for 8.9 \micron~as estimated from delete-$d$ jackknifing. The white contour is S/N$=3$ and the black contour shows S/N$=5$.}
    \label{fig:app_snr8p5}
\end{figure*}

\begin{figure*}
    \centering
    \includegraphics[width=0.9\textwidth]{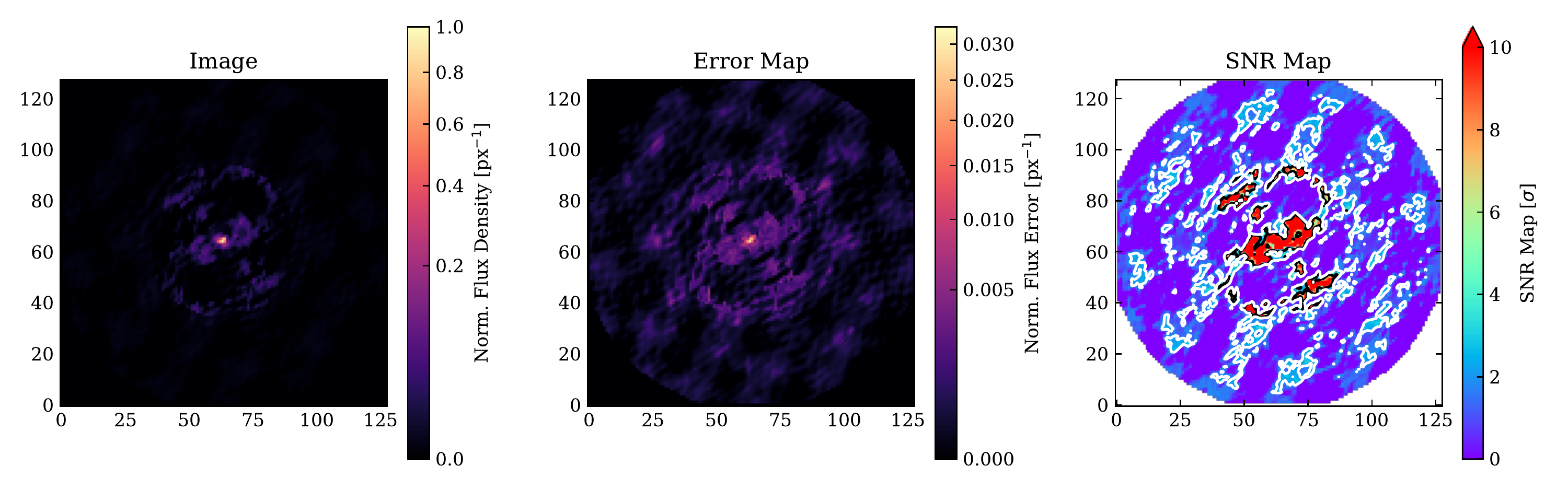}
    \caption{As Fig. \ref{fig:app_snr8p5}, but for 8.9 \micron.}
    \label{fig:app_snr8p9}
\end{figure*}

\begin{figure*}
    \centering
    \includegraphics[width=0.9\textwidth]{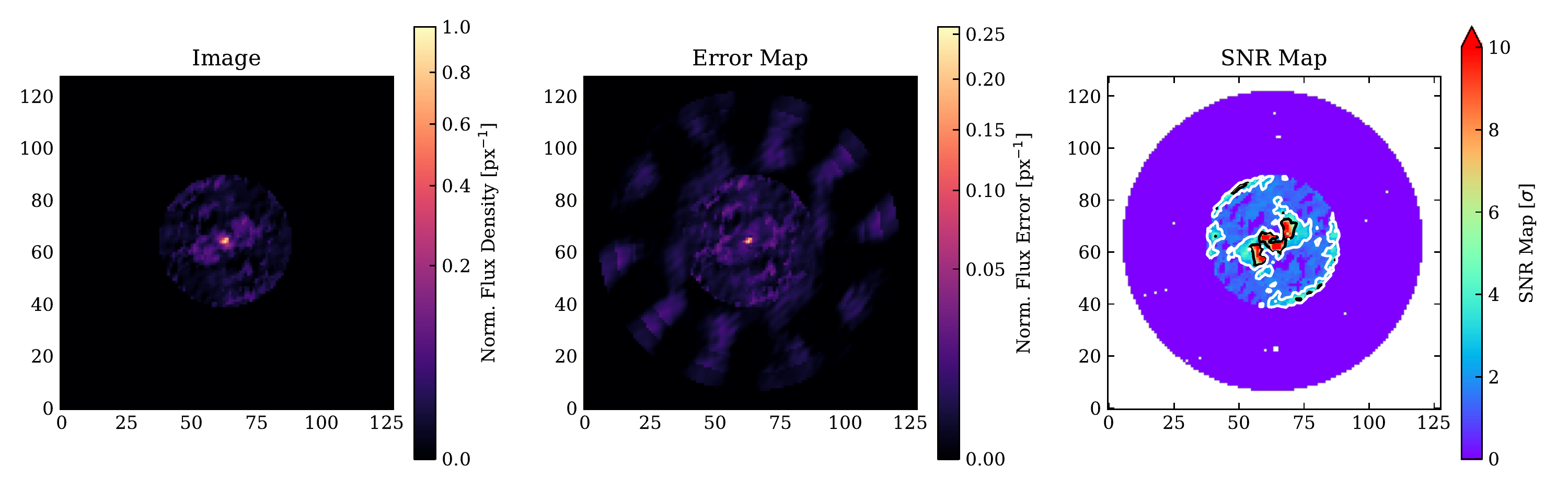}
    \caption{As Fig. \ref{fig:app_snr8p5}, but for 9.7 \micron.}
    \label{fig:app_snr9p7}
\end{figure*}

\begin{figure*}
    \centering
    \includegraphics[width=0.9\textwidth]{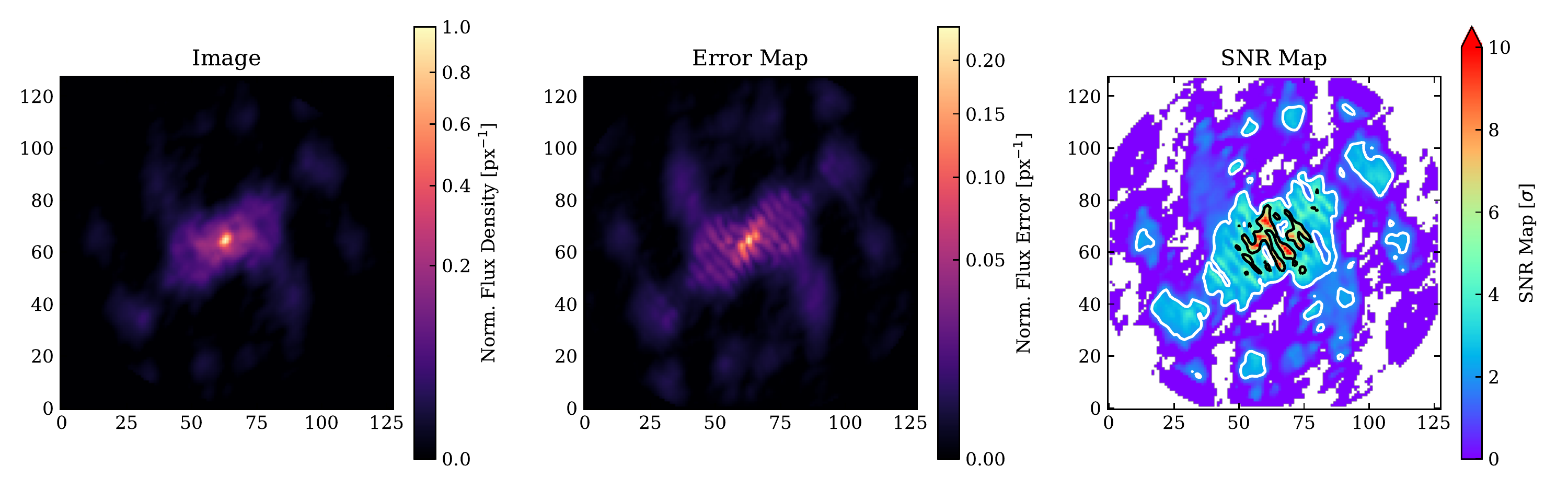}
    \caption{As Fig. \ref{fig:app_snr8p5}, but for 10.5 \micron.}
    \label{fig:app_snr10p5}
\end{figure*}

\begin{figure*}
    \centering
    \includegraphics[width=0.9\textwidth]{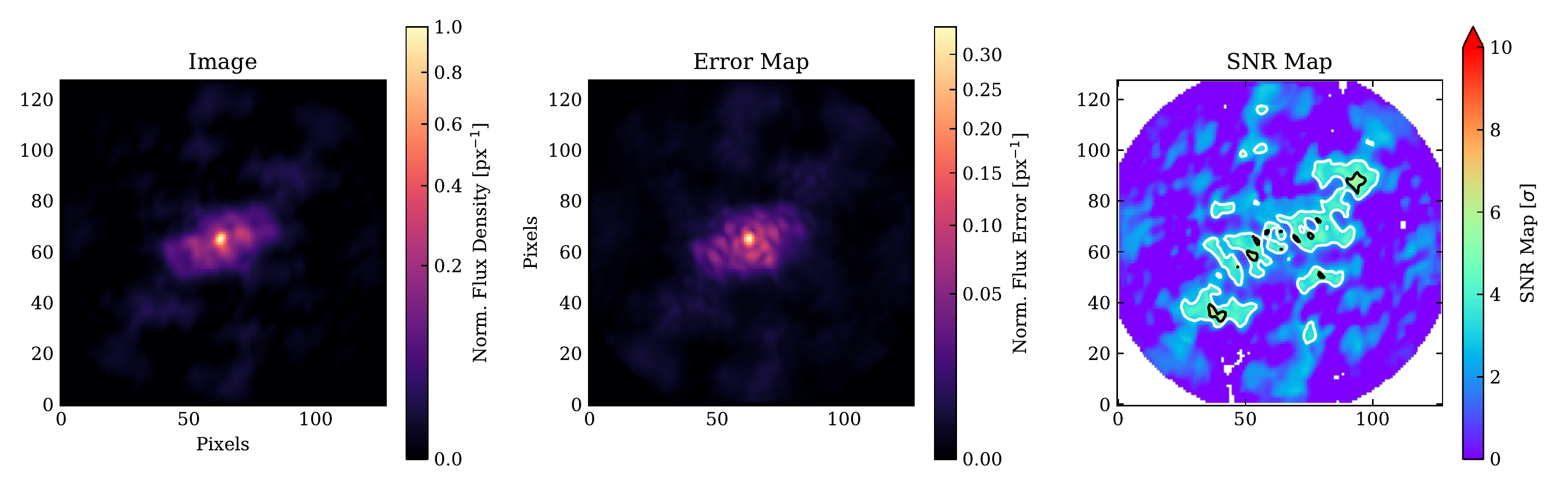}
    \caption{As Fig. \ref{fig:app_snr8p5}, but for 11.3 \micron.}
    \label{fig:app_snr11p3}
\end{figure*}

\begin{figure*}
    \centering
    \includegraphics[width=0.9\textwidth]{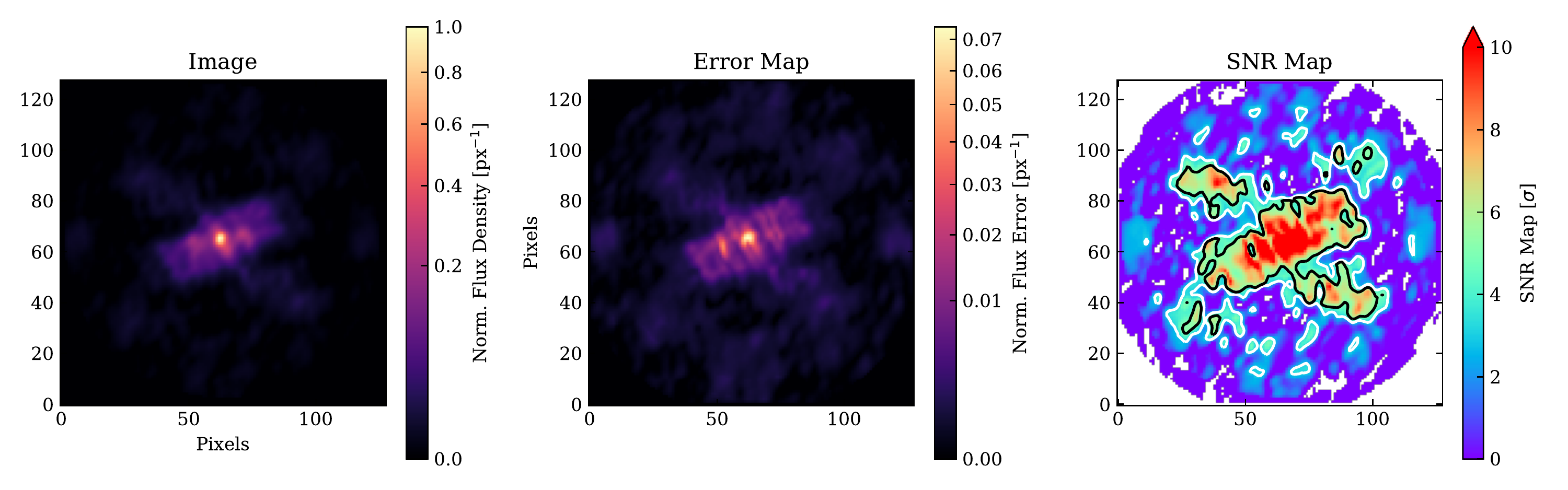}
    \caption{As Fig. \ref{fig:app_snr8p5} but for 12.0 \micron.}
    \label{fig:app_snr12p0}
\end{figure*}

\begin{figure*}
    \centering
    \includegraphics[width=0.9\textwidth]{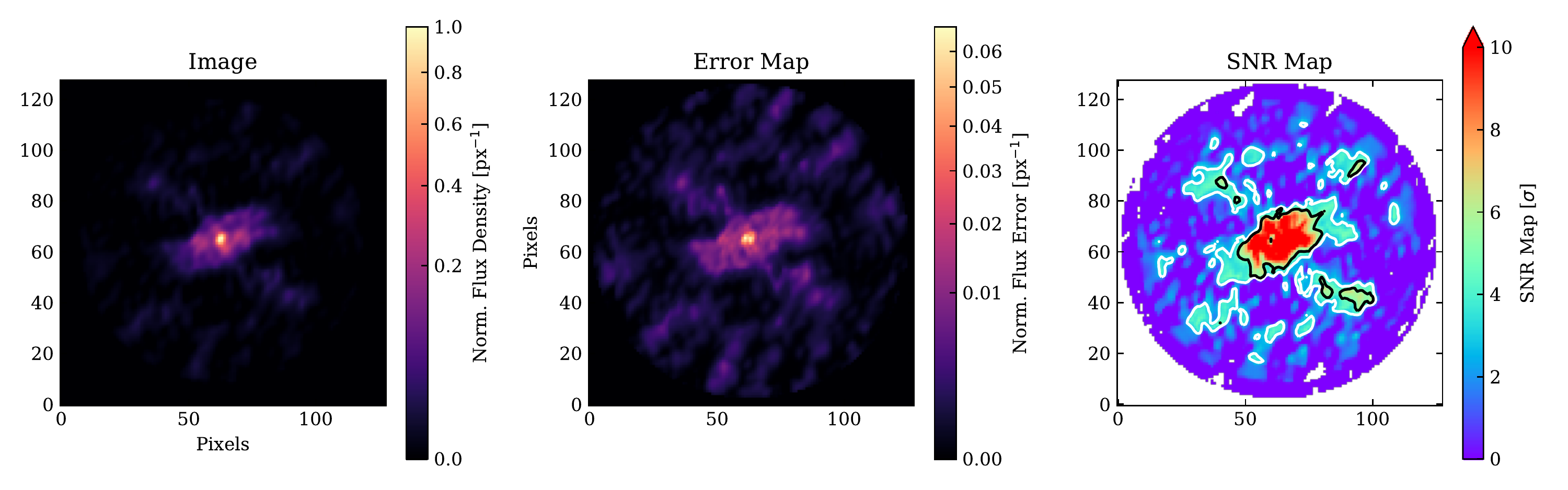}
    \caption{As Fig. \ref{fig:app_snr8p5}, but for 12.7 \micron.}
    \label{fig:app_snr12p7}
\end{figure*}

We performed delete-$d$ jackknifing \citep{shao1986} to estimate the errors present in our final images (see \S \ref{sec:imerrors}). We present the final images, the error maps, and the S/N maps in Figs. \ref{fig:app_snr8p5}-\ref{fig:app_snr12p7}. We use the S/N maps to determine which morphological features we trust. We perform an S/N cut of $\geq 3$ on the final images to $a)$ define where valid apertures can be located, and $b)$ determine the extent of large features.

\section{SKIRT model parameter variation}
\begin{figure*}
    \centering
    \includegraphics[width=0.75\textwidth]{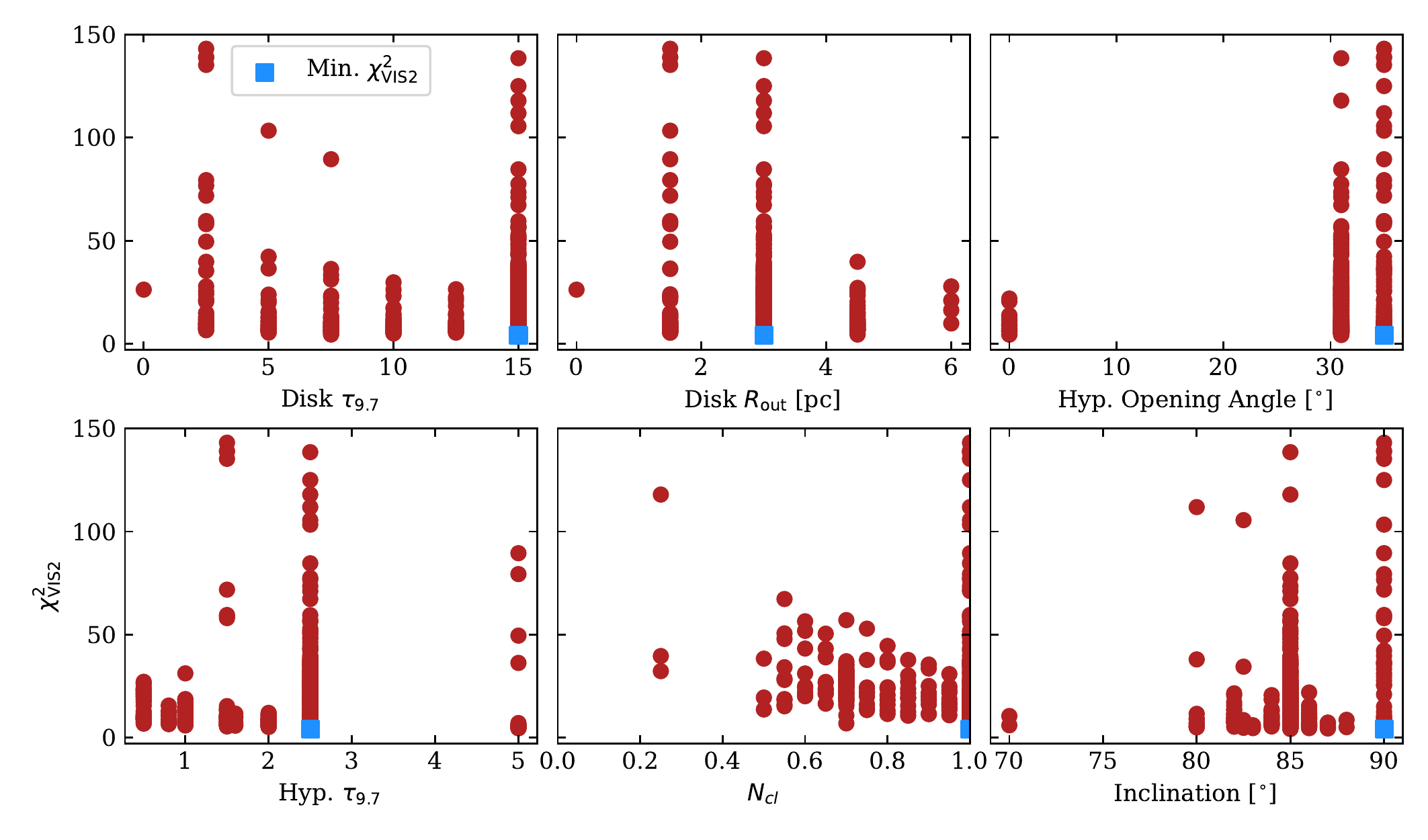}
    \includegraphics[width=0.75\textwidth]{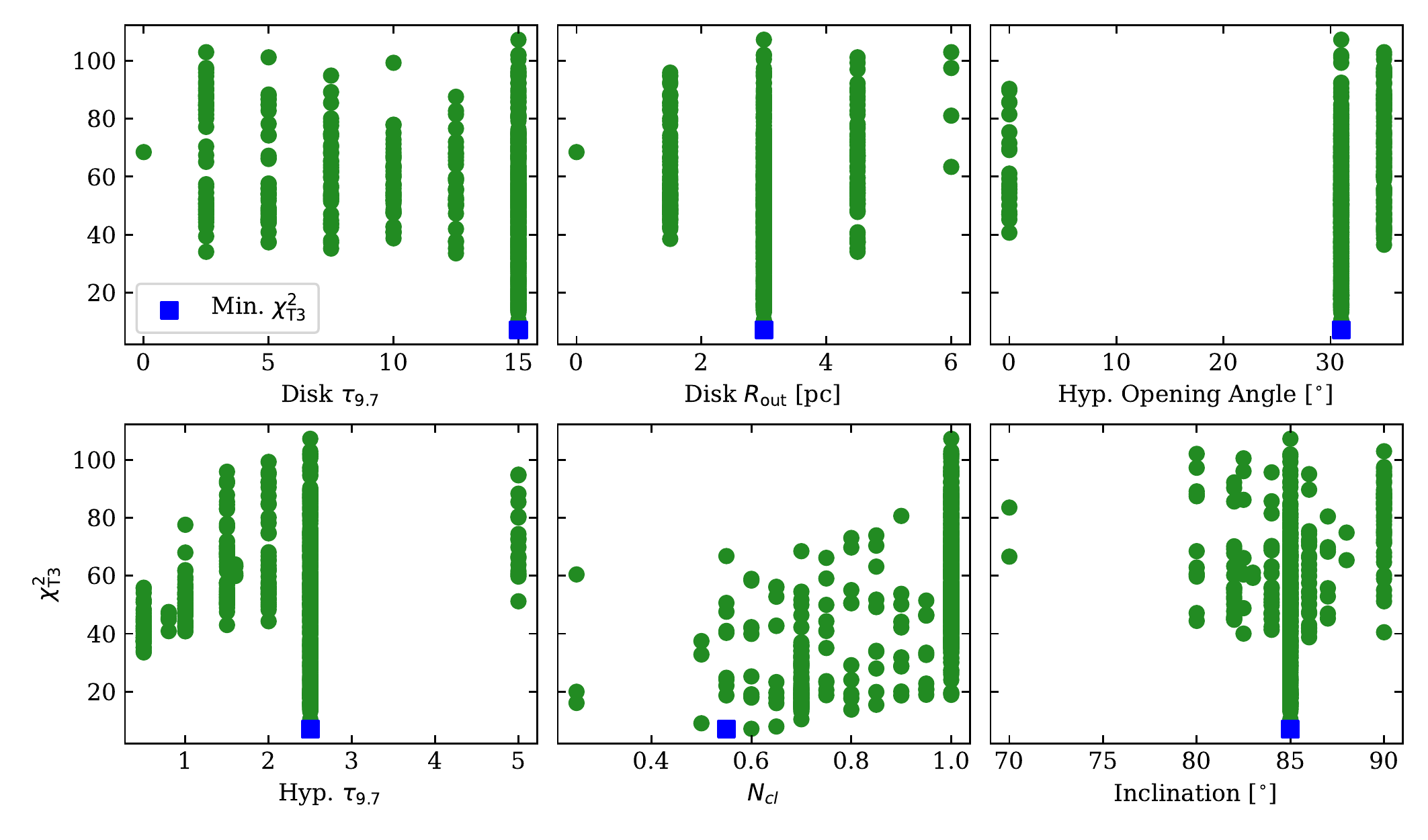}
    \includegraphics[width=0.75\textwidth]{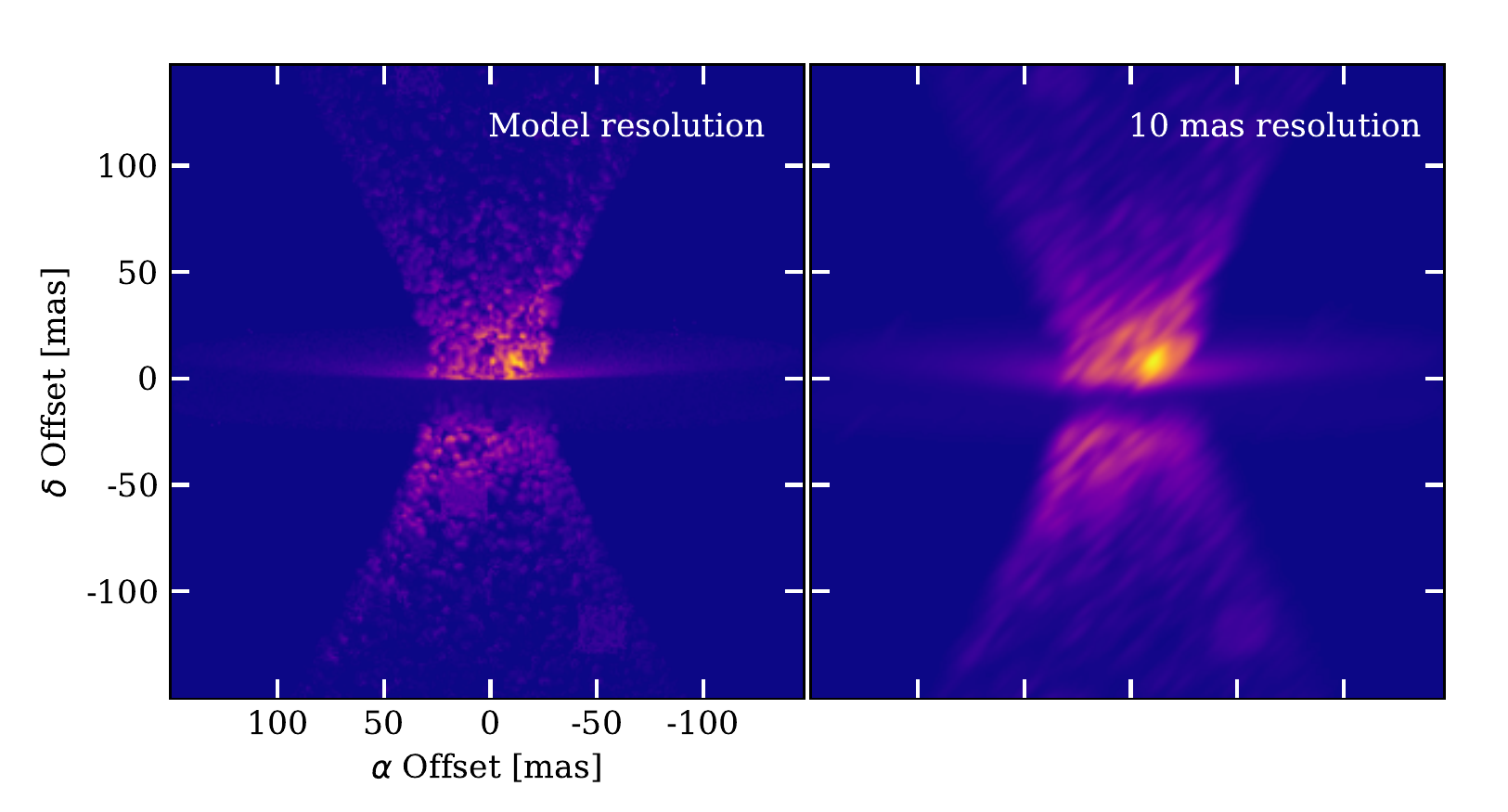}
    \caption{Comparisons via $\chi^2$ of simulated observables -- squared visbilities and closure phases -- to the MATISSE data for a range of disk+hyp model parameter values. In the \textit{top} six panels we show the squared visibility comparisons. In the \textit{middle} six panels we show the closure phase comparisons. In the \textit{bottom} panels we show the model with parameters favored by the $\chi^2$ comparison at both its native and 10 mas resolution.}
    \label{appfig:models_observables}
\end{figure*}

In Fig \ref{appfig:models_observables} we compare the extracted squared visibilities and closure phases to the observed values for a range of model parameters. The simulated squared visibilities and closure phases use the \uv-coverage of the MATISSE UTs. The parameters are the depth of the silicate feature in the disk and in the hyperboloid ($\tau_{9.7}$), the outer radius of the disk (Disk $R_{\rm out}$), the opening angle of the hyperboloid, the relative number of clumps ($N_{cl}$), and the inclination of the model ($i$ where 90$^{\circ}$ is edge-on). The comparisons place constraints on the system inclination ($i \approx 85^{\circ}$), the hyperboloid opening angle ($\theta_{\rm OA} \approx 30^{\circ}$), the disk Si feature depth ($\tau_{\rm Si, DSK} \approx 14$), and the outer radius of the disk ($r_{\rm out} \approx 3$ pc). The closure phases provide clearer constraints.

\begin{figure*}
    \centering
    \includegraphics[width=0.9\textwidth]{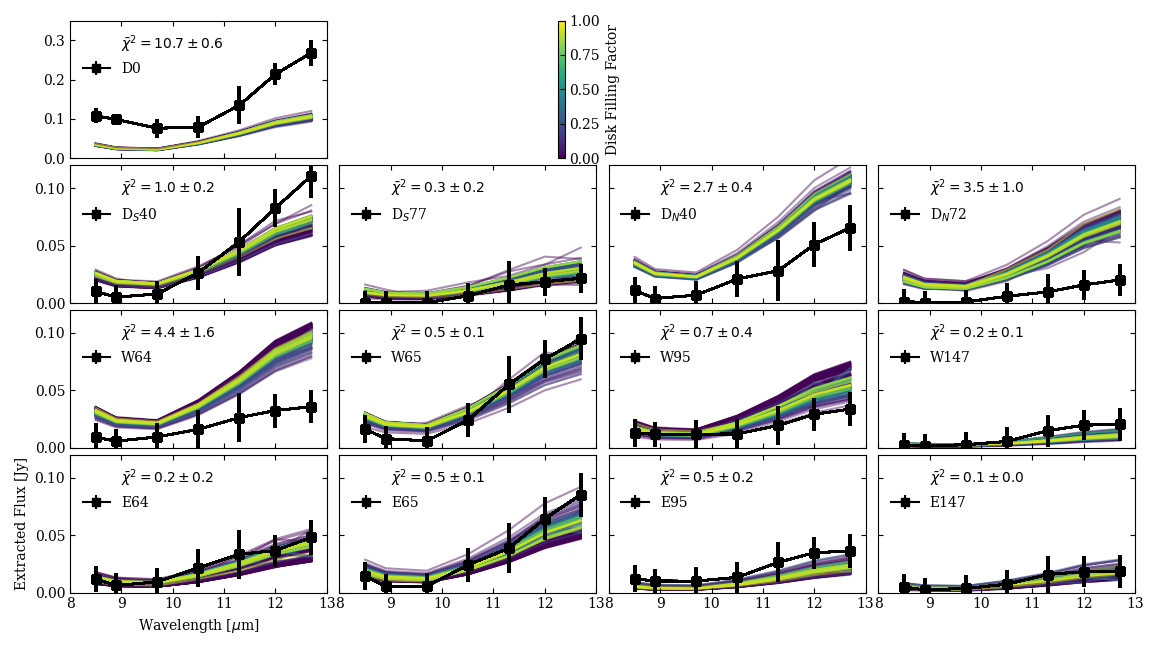}
    \caption{Comparisons of measured aperture-extracted spectra to those of disk+hyp models with disk filling factor varied. The displayed $\chi^2$ in each panel is the mean $\chi^2$ of all models and the ranges are given by the standard deviation of the model values. At large radii, the models agree well with observations, but the unresolved central flux is under-represented in the models.}
    \label{appfig:models_ff}
\end{figure*}

In Fig. \ref{appfig:models_ff} we show a comparison via $\chi^2$ between the model spectra and observed spectra in each of our 13 apertures defined in \S \ref{sec:temperature}. We see that flux in the central apertures is under-represented in the models, indicating that modifications to the disk component (of e.g., clumpiness or thickness) may be necessary.

\end{appendix}

\end{document}